\documentclass[traditabstract]{aa} 
\usepackage{graphicx,natbib,lscape,aalongtable}
\usepackage{txfonts}
\bibpunct{(}{)}{;}{a}{}{,}

\begin{document}
   \title{Elemental abundances and classification of CEMP stars
           \thanks{Tables \ref{fotomet}, \ref{temps}, \ref{paramDMA-SAT}, \ref{medDMA},
           \ref{medSAT}, \ref{compar}, \ref{alpha}, \ref{ironpeak}, \ref{heavy1},
           and \ref{heavy2} are only available in electronic form at http://www.edpsciences.org.}
	   \fnmsep\thanks{Tables \ref{DMAiron}, \ref{linesSAT}, and \ref{abunDMA} is only available 
           in electronic form 
	   at the CDS via anonymous ftp to cdsarc.u-strasbg.fr/Abstract.html (130.79.128.5)
	   or via http://cdsweb.u-strasbg.fr/cgi-bin/qcat?J/A+A/}}


   \author{D. M. Allen,
          \inst{1,2}
	  S. G. Ryan,
	  \inst{3,4}
	  S. Rossi,
	  \inst{1}
          T. C. Beers
	  \inst{5,6}
          \and
          S. A. Tsangarides
	  \inst{4}
	  }

   \institute{Instituto de Astronomia, Geof\'\i sica e Ci\^encias Atmosf\'ericas,
              Universidade de S\~ao Paulo, \\
	      Rua do Mat\~ao 1226, 05508-900 S\~ao Paulo, Brazil \\
              \email{dimallen@astro.iag.usp.br,rossi@astro.iag.usp.br}
	 \and
	     current address: Escola de Artes, Ci\^encias e Humanidades, USP,
             Av. Arlindo Bettio, 1000, 03828-000, \\
	     S\~ao Paulo, SP, Brazil
         \and
             Centre for Astrophysics Research, STRI and School of Physics, 
	     Astronomy and Mathematics, \\
	     University of Hertfordshire, Hatfield, UK \\
             \email{s.g.ryan@herts.ac.uk}
         \and
             previous address: The Open University, \\
	     Walton Hall, Milton Keynes, UK \\
	     \email{tsangarides.stelios@gmail.com}
        \and
             National Optical Astronomy Observatory,\\
             950 N. Cherry Avenue, Tucson, AZ 85719, USA 
             \email{beers@noao.edu}
         \and
	     previous address: Dept. of Physics \& Astronomy and JINA: 
	     Joint Institute for Nuclear Astrophysics, \\
	     Michigan State University, E. Lansing, MI 48824, USA
             }

   \date{Received 2010; accepted}

\authorrunning{Allen et al.}

\titlerunning{Elemental abundances and classification of CEMP stars}

 
  \abstract
   {We present a detailed study of Carbon-Enhanced Metal-Poor (CEMP) stars, 
    based on high-resolution spectroscopic observations of a sample of 18 stars.
    The stellar spectra for this sample were obtained at the 4.2m 
    William Herschel Telescope (WHT) in 2001 and 2002, using the Utrecht Echelle 
    Spectrograph (UES), at a resolving power $R \sim$ 52000 and S/N $\sim$ 40, covering the 
    wavelength range $\lambda\lambda$3700-5700 \AA. The atmospheric parameters 
    determined for this sample indicate temperatures ranging from 4750~K to 7100~K, 
    $\log g$ from 1.5 to 4.3, and metallicities $-$3.0 $\le$ [Fe/H] $\le$ $-$1.7. 
    Elemental abundances for C, Na, Mg, Sc, Ti, Cr, Cu, Zn, Sr, Y, Zr, Ba, La, Ce, 
    Nd, Sm, Eu, Gd, Dy are determined. Abundances for an additional 109 stars were taken from the literature 
    and combined with the data of our sample.  The literature sample reveals a lack of reliable abundance estimates for species that
    might be associated with the r-process elements for about 
    67\% of CEMP stars, preventing a complete understanding of this class of stars, 
    since [Ba/Eu] ratios are used to classify them. 
    Although eight stars in
    our observed sample are also found in the literature sample, 
    Eu abundances or limits are determined for four of these stars for the first time.
    From the observed correlations between C, Ba, and Eu, we argue that the CEMP-r/s 
    class has the same astronomical origin as CEMP-s stars, 
    highlighting the need for a more complete understanding of Eu production.}

   \keywords{stars: abundances -- stars: AGB and post-AGB -- stars:
population II -- stars: metal-poor
               }

   \maketitle
%

\section{Introduction}

Over the course of the past few decades, several groups have actively
studied stars identified as very metal poor by two major objective-prism
surveys, the HK Survey of Beers and colleagues \citep{beers85,beers92}
and the Hamburg/ESO survey of Christlieb and colleagues
\citep{christlieb08}. A large fraction, at least 20\%, of stars with
metallicities [Fe/H] $< -2.0$ have been shown to exhibit large
over-abundances of carbon relative to iron, [C/Fe] $>$ +1.0 
\citep{rossi99,lucatello06,marsteller09,carollo12}. 
The fraction of so-called carbon-enhanced metal-poor (CEMP) stars rises
to 30\% for [Fe/H] $< -3.0$, 40\% for [Fe/H] $< -3.5$, and 75\% for
[Fe/H] $< -4.0$ \citep{christ02,beers05,frebel05,norris07,carollo12}.
The CEMP fraction for ultra metal-poor stars with [Fe/H] $< -4.0$ was
100\% until the recent report by \citet{caffau11} of a new hyper
metal-poor star (SDSS~J102915+172927, with [Fe/H] $= -5.0$) that differs
from the other three known ultra and hyper metal-poor stars in that it
does not exhibit carbon enhancement. 

The diversity of chemical compositions seen among the CEMP stars,
together with their very low metallicities, provides insights into the
early stages of Galactic star formation, stellar evolution, and galaxy
formation. Several scenarios have been proposed to explain the likely
multiple origins of the high carbon enhancements shown by these stars,
but uncertainties remain. Observations of the neutron-capture elements
(those produced by the s-process and r-process) can help shed light on
these questions, given that their production is related to different
timescales, and almost certainly different astrophysical sites. In this
regard, it is of importance that \citet{hansen11} have recently reported
the results of a four-year radial velocity monitoring project for
moderately and highly r-process-enhanced metal-poor stars 
\citep[classes r-I and r-II according to the nomenclature of][]{beers05}, 
and demonstrated that the binary fraction of these stars is no different
than found for other low-metallicity halo stars (about 20\%). In
comparison, s-process-enhanced CEMP stars possess elevated binary
fractions, possibly 100\%, \citep{lucatello05}. This result immediately
suggests that the r-process enhancement phenomenon does not apparently
require pollution from a companion star, and underscores the likelihood
that the astrophysical sites of the r-process and s-process are
decoupled from one another.

Barium and europium are often taken as representatives of the s- and the
r-processes, respectively, according to the dominance of these processes
on their production in Solar System material. According to
\citet{arlandini99}, in the Solar System, the main component of the
s-process contributes $\approx$ 82\% of the Ba, whereas $\approx$ 18\%
is due to the r-process. By way of contrast, while 6\% of Eu in
solar-composition material results from the s-process, the r-process
accounts for the production of essentially all the rest (94\%). Barium
abundances are easily measured for very metal-poor stars, since its
strongest lines are unsaturated, which is often not the case for stars
with higher metallicities. Although the lines of Eu are weaker than
those of Ba, Eu abundances nevertheless are often readily measurable in
optical spectra of metal-poor stars. Hence, Ba and Eu are useful and
accessible indicators of the s- and r-processes.

The r-process is believed to occur in the final evolutionary stages of
massive stars ($M$ $>$ 8 M$\sb \odot$), whereas the main component of
the s-process is believed to occur in thermally-pulsing asymptotic
giant-branch (TP-AGB) stars of low (1-3 M$\sb \odot$) to intermediate
(4-8 M$\sb \odot$) masses \citep[e.g., ][]{herwig05}. As the timescale
for stars to reach the SN II (type II supernova) stage 
($t$ $<$ 10$\sp 8$ yr) is less than
that for stars to reach the AGB phase, it is usual to investigate the
s-process contribution by analyzing the [Ba/Eu] abundance ratio. A value
of [Ba/Eu] $\sim -$0.70, found in metal-poor stars, indicates Ba
production only by the r-process \citep{mg01}. When the s-process begins
to produce Ba, an increase in its abundance relative to Eu can be seen,
ultimately reaching values compatible with the solar [Ba/Eu] ratio
\citep{travaglio99}. 

\citet{beers05} have used the Ba and Eu abundance
signatures to classify CEMP stars into several broad categories: CEMP-s,
stars that exhibit large over-abundances of s-process elements, with
[Ba/Fe] $>$ +1 and [Ba/Eu] $>$ +0.5, and account for some 80\% of all
studied CEMP stars according to \citet{aoki07}; CEMP-r, stars exhibiting
over-abundances of r-process elements and [Ba/Eu] $<$ 0 \citep{mcwill95,
norris97,sneden03, honda04,barklem05,cohen06}; CEMP-r/s, stars with 0
$<$ [Ba/Eu] $\lesssim$ +0.5, and exhibiting both r- and s-process
enhancements \citep{hill00, wanajo06}; and CEMP-no, stars that exhibit
no over-abundances of neutron-capture elements \citep[{[Ba/Fe]} $<$ 0;
e.g.,][]{aoki07,ito09}.

There has been some discussion in the literature concerning the
relationship of the CEMP-r/s class to the CEMP-s class. \citet{tsang05}
pointed out that the [Ba/Eu] ratios for some stars classified previously
as CEMP-s are similar to those for stars classified as CEMP-r/s, and
called into question whether the hybrid r/s classification is strictly
necessary. He also speculated that the progenitors of the CEMP-s and
CEMP-r/s class may even be one and the same (TP-AGB stars).
Unfortunately, for many known CEMP stars, measurements of Eu abundances
are not available, and instead Ba is sometimes used as an r-process
indicator, even though it is produced in both the r- and s-processes.
For further discussion of these issues, see the recent paper by
\citet{bisterzo10}.

In this paper we seek a better understanding of the possible differences
between stars classified as CEMP-s and CEMP-r/s, by expanding the
numbers of CEMP stars with available Eu abundance measurements. We also
reconsider the interpretation of previously published high-resolution
abundance determinations for CEMP stars. The paper is organized as
follows. Section \ref{param} presents the data and describes the
determination of the stellar atmospheric parameters. Section
\ref{abundt} describes the abundance and uncertainty determinations. In
Section \ref{disc} our results are discussed, as well as those obtained
from inspection of data from the literature. Our conclusions are drawn
in Section \ref{concl}.					    


\section{Observations and stellar parameters}\label{param}

Twenty-six candidate CEMP stars, listed in Table \ref{fotomet}, were
selected from the HK survey of \citet{beers92}, and from more recent
unpublished follow-up medium-resolution extensions of that programme,
according to two criteria: An HK-survey metallicity estimate [Fe/H]
$\lesssim -$ 1.5, and a carbon-enhancement index CEN $\gtrsim$ 0.5,
implying a CH G-band strength more than 1.5x the median value at that
temperature and metallicity. The CEN index is defined as
(GP$_*$~-~GP$\rm _M$) /GP$\rm _M$, where GP is the \citet[][their Table
1]{beers92} CH G-band index, and the subscripts ``*'' and ``M'' refer to
a particular star and the median for stars of that temperature and
metallicity. The notes column identifies stars which our analysis
ultimately showed were not genuine CEMP stars (Section~3.1).

Twenty-three stellar spectra were observed by 
SAT\footnote{Stelios A. Tsangarides} and 
SGR\footnote{Sean G. Ryan} at the 4.2m
William Herschel Telescope (WHT) in August 2001 and April 2002, using
the Utrecht Echelle Spectrograph (UES) , with a resolving power of R
$\sim$ 52000 and S/N $\sim$ 40 per 0.026 $\rm \AA$ pixel. The spectra
cover the wavelength range $\lambda\lambda$3500 - 5700 \AA. Three
additional high-resolution spectra were obtained by SGR and John Norris
on the 3.9m Anglo-Australian Telescope in September 2000, using the UCL
Echelle Spectrograph (UCLES). The spectra were reduced by SAT with
IRAF\footnote{IRAF is distributed by the National Optical Astronomy
Observatory, which is operated by the Association of Universities for
Research in Astronomy (AURA) under cooperative agreement with the US
National Science Foundation.}, using standard procedures. Because of the
large number of spectra taken, the reduced data were analysed in two
subgroups, one by DMA\footnote{Dinah M. Allen} (12 stars) 
and the other by SAT (14 stars). The
approaches were similar in concept, although they differed slightly in
the details. 

Photometric {\it UBVRI} observations of stars from the SAT subsample
(see Table \ref{fotomet}) were obtained by TCB\footnote{Timothy C. Beers} on the
Wisconsin-Indiana-Yale-NOAO (WIYN) Observatory 0.9m telescope on Kitt
Peak, and at the ESO/Danish 1.5m telescope at La Silla (Chile), and by
TCB and Don Terndrup on the MDM Observatory 2.5m telescope, also on Kitt
Peak \citep{beers94,beers07}. Near-IR {\it JHK} photometry was taken for
the 26 objects from the interim data release of the 2MASS point source
catalogue \citep[2MASS 2003; ][]{skrutskie97}. 

Calculation of even a single model atmosphere is a computationally
intensive task. Stellar parameters, such as effective temperature
($T_{\rm eff}$), surface gravity ($\log g$), and metallicity ([Fe/H]),
are usually required a-priori. However, these are quantities to be
determined using the model atmosphere during the analysis; thus, we can
only arrive at an appropriate set of parameters and model atmosphere by
iteration. This procedure is impractical if a fresh model is to be
obtained at every step, given the computational demands of its
calculation. Hence, most groups evaluate a grid of several model
atmospheres to begin with, in which each model possesses a different set
of input atmospheric parameters. One then interpolates between models of
this grid, in order to more quickly compute new models at a desired set
of atmospheric parameters. Inherent in this interpolation is the
assumption that the physics of each grid model can be scaled to nearby
atmospheric parameters.

\onltab{1}{
\begin{table*}
{\scriptsize
\caption{Colours and magnitudes of our program stars before dereddening. The
last four columns are the Galactic coordinates, the S/N ratio of the high-resolution spectra around $\lambda$4500, and (for some stars) notes. 
A suffix ``X'' indicates that the value was not used.
}
\label{fotomet} 
\centering
\setlength\tabcolsep{3pt}
\begin{tabular}{cccllcrrrcccrrrl}
\noalign{\smallskip}
\hline \hline
Star & $R.A.$ & $DEC$ & V & E(B-V) & $A_{\rm V}$ & (B-V) & (V-R) & (V-I) & J & H & K$_s$ & l & b & S/N & Notes\\
\hline
\noalign{\smallskip}
&&&&\multicolumn{4}{c}{DMA subsample} &&&&&&& \\
\hline
\noalign{\smallskip}
BS 15621-047 & 10 09 35.7  & +25 59 32   & 13.735 & 0.058 & 0.180(160) & 0.880 &  0.504 &  0.992 & 11.955(21) & 11.434(19) & 11.277(19) & 205.3 & +54.0 &  39 \\
BS 16033-081 & 13 19 12.4  & +22 27 57   & 13.347 & 0.067 & 0.210(160) & 0.782 &  0.496 &  0.988 & 11.665(25) & 11.158(28) & 11.070(24) & 357.9 & +82.2 &  47 \\
BS 16077-077 & 11 55 16.6  & +28 21 02   & 13.70  & 0.029 & 0.092(150) & 0.610 &   ...  &    ... & 10.999(22) & 10.710(22) & 10.652(18) & 204.1 & +77.5 &  74 \\
BS 16082-129 & 13 47 11.5  & +28 57 46   & 13.483 & 0.051 & 0.160(170) & 0.731 &  0.470 &  0.942 & 11.855(22) & 11.402(21) & 11.309(16) &  44.7 & +77.6 &  45 \\
BS 16543-097 & 13 26 28.3  & +21 38 42   & 12.570 & 0.032 & 0.100(180) & 0.691 &  0.457 &  0.913 & 11.007(23) & 10.573(25) & 10.456(18) &   0.3 & +80.3 &  40 \\
BS 16929-005 & 13 03 29.4  & +33 51 06   & 13.610 & 0.032 & 0.100(150) & 0.618 &  0.402 &  0.844 & 12.172(22) & 11.751(20) & 11.669(20) & 102.6 & +82.8 &  45 \\
CS 22949-008 & 23 16 58.5  & -03 20 46   & 14.168 & 0.022 & 0.070(150) & 0.494 &  0.310 &  0.623 & 13.123(23) & 12.822(26) & 12.788(30) &  75.3 & -57.2 &  37 \\
CS 29503-010 & 00 04 55.4  & -24 24 19   & 13.737 & 0.032 & 0.100(160) & 0.472 &  0.307 &  0.609 & 12.726(24) & 12.468(26) & 12.433(26) &  44.9 & -79.2 &  51 \\
CS 29512-073 & 22 28 36.4  & -12 24 42   & 14.137 & 0.054 & 0.170(150) & 0.568 &  0.377 &  0.753 & 12.902(24) & 12.558(25) & 12.513(30) &  49.6 & -53.5 &  37 \\
CS 29526-110 & 03 27 43.6  & -23 00 31   & 13.352 & 0.051 & 0.160(150) & 0.356 &  0.236 &  0.518 & 12.587(26) & 12.367(25) & 12.367(24) & 215.1 & -54.5 &  37 \\
CS 29528-028 & 02 28 26.5  & -21 33 00   & 14.510 & 0.032 & 0.100(150) & 0.352 &  0.218 &  0.495 & 13.755(28) & 13.535(29) & 13.537(40) & 204.0 & -67.2 &  30 \\
CS 31070-073 & 00 08 51.7  & +05 26 16   & 14.558 & 0.022 & 0.070(150) & 0.449 &  0.315 &  0.655 & 13.535(27) & 13.258(31) & 13.203(29) & 103.8 & -55.8 &  34 \\
\hline\hline
\noalign{\smallskip}
&&&&\multicolumn{4}{c}{SAT subsample} &&&&& \\
\noalign{\smallskip}
\hline
BS 16080-175 & 16 51 08.8  & +57 12 28   & 14.35  & 0.028 &   ...      & 0.450 &  0.364 &  0.678 & 13.235(23) & 12.977(24) & 12.960(30) &  86.0 & +38.7 &  37 \\
BS 17436-058 & 13 36 46.0  & +45 03 08	 & 13.83  & 0.016 &   ...      & 0.740 &  0.429 &  0.856 & 12.414(19) & 12.024(21) & 11.954(23) &  99.1 & +69.9 &  36 \\
CS 22183-015 &  01 00 53.0 & -02 28 20   & 13.74  & 0.041 &   ...      & 0.680 &  0.409 &  0.798 & 12.405(23) & 12.063(22) & 12.027(24) & 128.6 & -65.2 &  49 \\
CS 22887-048 & 22 46 44.6  & -11 08 42	 & 12.87  &	  &   ...      & 0.390 &  0.282 &  0.560 & 11.965(24) & 11.743(25) & 11.674(25) & 055.5 & -56.6 &  47 \\
CS 22898-027 & 21 05 45.1  & -18 36 57	 & 12.76  & 0.063 &   ...      & 0.500 &  0.284 &  0.666 & 11.719(22) & 11.503(26) & 11.421(25) & 029.7 & -37.7 &  45 \\
CS 29502-092 & 22 22 35.9  & -01 38 27	 & 11.87  & 0.099 &   ...      & 0.770 &  0.505 &  1.008 & 10.177(27) &  9.713(22) &  9.601(19) & 062.1 & -46.2 & 100 \\
\hline
\noalign{\smallskip}
BS 16090-048 &  14 31 27.5 & +52 03 45	 & 14.10  & 0.011 &   ...      & 0.457 &  0.286 &  0.558 & 13.105(23) & 12.922(37) & 12.850(25) & 92.8  & +58.8 & 26 \\
CS 22171-009 & 01 57 51.5  & -08 46 34	 & 13.37  & 0.026 &   ...      & 0.420 &  0.279 & 0.593  & 12.331(24) & 12.074(23) & 12.021(28) & 166.6 & -65.8 & 37 \\
CS 22174-007 & 01 14 06.5  & -11 02 35	 & 12.40  & 0.027 &   ...      & 0.721 &  0.448 & 0.900	 & 10.404(26) & 10.404(22) & 10.318(21) & 142.3 & -73.1 & 41 \\
\hline
\noalign{\smallskip}
BS 16086-022 & 15 06 27.7  & +60 18 21   & 13.281 & ...   &    ...     & 0.643 &  0.424 &  0.800 & 11.914(23) & 11.521(21) & 11.461(18) &  97.7 & +49.7 &  47 & DQ WD?\\
BS 16088-104 & 16 50 28.5  & +47 12 42   & 14.047 & ...   &    ...     & 0.676 &  0.458 &  0.792 & 12.812(23) & 12.486(23) & 12.436(27) &  73.2 & +39.7 &  38 & not CEMP\\
BS 16542-052 & 16 27 43.6  & +21 03 51   & 12.76  & 0.05  &    ...     & 0.870 &  0.553 &  1.01  & 11.063(22) & 10.625(20) & 10.533(18) &  38.2 & +40.6 &  47 & emission?\\
BS 17451-031 & 15 37 20.5  & +63 47 05	 & ...	  & 0.016 &   ...      & 0.711X&   ...  &   ...  &   ...      &   ...      &   ...      &  98.6 & +44.7 &  27 & colour doubtful\\
BS 17451-031 & 15 37 20.5  & +63 47 05   &	  & 0.016 &   ...      & 0.449 &   ...  &   ...  &   ...      &   ...      &   ...      &  98.6 & +44.7 &  27 & not CEMP\\
CS 22169-002 & 04 00 38.0  & -15 09 19	 & 13.33  & ...   &   ...      & 0.480 &  0.374 &  0.743 & 12.106(29) & 11.723(23) & 11.660(26) & 207.6 & -44.6 &  53 & DQ WD?\\
\noalign{\smallskip}
\hline
\end{tabular}
}
\end{table*}
}

\begin{table}
\caption{Iron lines for the DMA subsample (first four lines only; the full table is only available on CDS). Equivalent widths are in m\AA .} 
\label{DMAiron}
\centering
\setlength\tabcolsep{3pt}
\begin{tabular}{lrrrrrrrr}
\noalign{\smallskip}
\hline \hline
ion & $\lambda$ ($\rm \AA$) & $\chi_{ex}$ (eV) & $\log gf$ & Ref. & 1  & 2 & 3 & 4 \\
\hline
\noalign{\smallskip}
Fe I  & 4001.661 &   2.176& -1.877 & 1 &    ... &    ... &    ... &    ... \\
Fe I  & 4005.242 &   1.557& -0.610 & 1 &    ... &    ... &  85.60 &    ... \\
Fe I  & 4009.713 &   3.65 & -1.203 & 1 &    ... &    ... &  36.50 &  44.60 \\
Fe I  & 4014.531 &   3.48 & -0.195 & 1 &  70.00 &  54.70 &  34.80 &    ... \\
\noalign{\smallskip}
\hline
\end{tabular}

\tablefoot{Stars in columns 6 to 17: (1) BS~15621-047 (16;11); (2) BS~16033-081 (86;18); (3) BS~16077-077 (124;20); (4) BS~16082-129 (88;14);
(5) BS~16543-097 (100;20); (6) BS~16929-005 (114;10); (7) CS 22949-008 (52;6); (8) CS 29503-010 (92;17); (9) CS~29512-073 (100;17); 
(10) CS~29526-110 (27;6); (11) CS~29528-028 (12;3); (12) CS~31070-073 (37;9). 
Numbers in parenthesis after the star name are the number of lines of Fe I and Fe II (FeI;FeII). References of the sources of $\log gf$ 
are shown in column 5.
}

\tablebib{
(1) NIST; (2) \citet{fw06}
}
\end{table}

\onltab{3}{
\begin{table*}
{\footnotesize
\caption{Temperatures derived from photometric data by using the final metallicity found for 
each star. For the DMA subsample, the first and second lines show the results for stellar parameters using  
[FeI/H] and [FeII/H], respectively. For the SAT subsample only one [Fe/H] value is given 
since ionisation balance was forced in determining $\log g$. 
A suffix ``X'' indicates that the value was not used.
} 
\label{temps} 
\centering
\setlength\tabcolsep{3pt}
\begin{tabular}{lrrrrrrr}
\noalign{\smallskip}
\hline \hline
Star & [Fe/H] & $T_{BV}$ & $T_{VR}$ & $T_{VI}$ & $T_{VK}$ & $T_{JK}$ & $T_{JH}$ \\
\hline
\noalign{\smallskip}
&&\multicolumn{3}{c}{DMA subsample} &&& \\
\hline
\noalign{\smallskip}
BS 15621-047 & -2.29 & 4730 & 4860 & 4870 & 4770 & 4620 & 4740 \\
             & -2.35 & 4730 & 4870 & 4860 & 4780 & 4620 & 4740 \\
\noalign{\smallskip}
BS 16033-081 & -2.20 & 5170 & 5020 & 5020 & 5140 & 5090 & 5040 \\
             & -2.07 & 5180 & 5020 & 5010 & 5130 & 5090 & 5060 \\
\noalign{\smallskip}
BS 16077-077 & -2.05 & 5440X& ...  & ...  & 4220X& 5890 & 5830 \\
             & -2.00 & 5450X& ...  & ...  & 4220X& 5890 & 5830 \\
\noalign{\smallskip}
BS 16082-129 & -2.65 & 5090 & 4980 & 5010 & 5100 & 5050 & 4970 \\
             & -2.59 & 5100 & 4980 & 5010 & 5090 & 5050 & 4970 \\
\noalign{\smallskip}
BS 16543-097 & -2.48 & 5170 & 5000 & 5030 & 5090 & 4990 & 5040 \\
             & -2.36 & 5170 & 5000 & 5020 & 5090 & 4990 & 5050 \\
\noalign{\smallskip}
BS 16929-005 & -3.15 & 5400 & 5190 & 5350 & 5360 & 5180 & 4990 \\
             & -3.02 & 5390 & 5190 & 5340 & 5350 & 5180 & 5020 \\
\noalign{\smallskip}
CS 22949-008 & -2.45 & 5740 & 5950 & 5920 & 5990 & 5960 & 5750 \\
             & -2.61 & 5740 & 5960 & 5920 & 6000 & 5960 & 5750 \\
\noalign{\smallskip}
CS 29503-010 & -1.69 & 5990 & 5950 & 5840 & 6260 & 6220 & 6050 \\
             & -1.70 & 5990 & 5950 & 5840 & 6260 & 6220 & 6050 \\
\noalign{\smallskip}
CS 29512-073 & -2.06 & 5680 & 5520 & 5460 & 5850 & 5750 & 5560 \\
             & -1.88 & 5690 & 5520 & 5460 & 5840 & 5750 & 5580 \\
\noalign{\smallskip}
CS 29526-110 & -2.19 & 6600 & 6390 & 6470 & 7100 & 6790 & 6320 \\
             & -2.29 & 6600 & 6390 & 6480 & 7110 & 6790 & 6300 \\
\noalign{\smallskip}
CS 29528-028 & -2.12 & 6500 & 6690 & 6790 & 6960 & 6840 & 6350 \\
             & -2.17 & 6500 & 6690 & 6790 & 6970 & 6840 & 6350 \\
\noalign{\smallskip}
CS 31070-073 & -2.55 & 6020 & 5750 & 5800 & 6200 & 5950 & 5820 \\
             & -2.64 & 6020 & 5750 & 5810 & 6210 & 5950 & 5800 \\
\hline\hline
\noalign{\smallskip}
&&\multicolumn{3}{c}{SAT subsample} &&& \\
\noalign{\smallskip}
\hline
BS 16080-175 & -1.36 & 5942 & 6329 & 6247 & ...  & 6328 & 5970 \\
\noalign{\smallskip}
BS 17436-058 & -1.90 & 4986 & 5246 & 5209 & ...  & 5416 & 5467 \\
\noalign{\smallskip}
CS 22183-015 & -2.85 & 5291 & 5416 & 5468 & ...  & 5600 & 5588 \\
\noalign{\smallskip}
CS 22887-048 & -1.70 & ...  & ...  & ...  & ...  & 6508 & 6581 \\
\noalign{\smallskip}
CS 22898-027 & -2.30 & 6039 & 6562 & ...  & ...  & 6477 & 6363 \\
\noalign{\smallskip}
CS 29502-092 & -3.05 & 5224 & 5064 & 5068 & ...  & 4978 & 5060 \\
\noalign{\smallskip}
\hline
\end{tabular}
}
\end{table*}
}

\onltab{4}{
\begin{table*}
{\footnotesize
\caption{Results for the stellar parameters, for the DMA (upper portion) 
and SAT (lower portion) subsamples. For the DMA subsample, $T_{\rm ex}$ 
is the excitation temperature, $M_{\rm bol}$ is the bolometric magnitude,
$L$ is the luminosity, $R$ is the radii, $M$ the mass, $\pi$ the parallax.
Uncertainties are shown in the second line.
For [Fe/H], the uncertainties are the standard error of the mean abundance, 
excluding the effects of errors on atmosphere parameters. SAT did not use 
isochrones, so did not determine values for  
 BC(V),  $M_{\rm V}$,  $M_{\rm bol}$,  $L$,  $R$, $M$ or $D$. 
 The last column indicates whether the star is binary or not, 
 according to \citet{tsang05}.}
\label{paramDMA-SAT} \centering
\setlength\tabcolsep{2pt}
\begin{tabular}{rrrrrrrrrrrrrcc}
\noalign{\smallskip}
\hline \hline
 $T_{\rm ex}$(K) & $\log g$ & [FeI/H] & [FeII/H] & $\xi$(km/s) & 
 BC(V) & $M_{\rm V}$ & $M_{\rm bol}$ & $L$/L$_\odot$ & $R$/R$_\odot$
& $M$/M$_\odot$ & $D$(pc) & [Ba/Eu] & Class & Var. \\
\hline
\noalign{\smallskip}
&&&&&\multicolumn{3}{c}{DMA subsample} &&&&&&& \\
\hline
&&&&&\multicolumn{3}{c}{BS 15621-047} &&&&&& \\
 4750 &  1.48 & -2.29 & -2.35 & 1.05 & -0.384 & -1.38 & -1.76 & 02.85 & 29.73 & 0.96 & 9700 & -0.49 & r & B \\
   70 &  0.09 &  0.07 &  0.09 & 0.05 &  0.025 &  0.21 &  0.20 & 76.37 &  2.97 & 0.09 & 1188 &&& \\
\hline
&&&&&\multicolumn{3}{c}{BS 16033-081} &&&&&& \\                  				      
 5020 &  1.84 & -2.20 & -2.07 & 1.2  & -0.297 & -0.61 & -0.91 & 82.92 & 17.94 & 0.80 & 5620 & -1.31 & r & B \\
   50 &  0.07 &  0.05 &  0.07 & 0.1  &  0.014 &  0.13 &  0.12 & 21.33 &  1.12 & 0.00 &  542 &&& \\
\hline
&&&&&\multicolumn{3}{c}{BS 16077-077} &&&&&& \\
 5900 &  3.19 & -2.05 & -2.00 & 1.00 & -0.164 &  1.76 &  1.59 & 18.29 &  4.11 & 0.95 & 2344 &  0.76 & s & \\
   50 &  0.08 &  0.05 &  0.09 & 0.20 &  0.007 &  0.20 &  0.20 &  3.45 &  0.41 & 0.09 &  279 &&& \\
\hline
&&&&&\multicolumn{3}{c}{BS 16082-129} &&&&&& \\
 5080 &  1.98 & -2.65 & -2.59 & 1.30 & -0.298 & -0.26 & -0.56 & 32.80 & 14.92 & 0.77 & 5200 &  ... & no?r? & \\
   50 &  0.11 &  0.06 &  0.08 & 0.05 &  0.014 &  0.27 &  0.27 & 33.05 &  1.89 & 0.13 &  773 &&& \\
\hline
&&&&&\multicolumn{3}{c}{BS 16543-097} &&&&&& \\
 5220 &  2.54 & -2.48 & -2.36 & 1.85 & -0.264 &  1.00 &  0.73 & 40.41 &  7.80 & 0.77 & 1970 & -0.49 & r & \\
   50 &  0.10 &  0.06 &  0.08 & 0.10 &  0.012 &  0.16 &  0.16 &  6.10 &  0.61 & 0.05 &  225 &&& \\
\hline
&&&&&\multicolumn{3}{c}{BS 16929-005} &&&&&& \\
 5250 &  3.10 & -3.15 & -3.02 & 0.90 & -0.288 &  2.40 &  2.11 & 11.39 &  4.09 & 0.77 & 1670 & -0.74 & r & \\
   50 &  0.09 &  0.06 &  0.09 & 0.10 &  0.010 &  0.22 &  0.22 &  2.29 &  0.42 & 0.10 &  207 &&& \\
\hline
&&&&&\multicolumn{3}{c}{CS 22949-008} &&&&&& \\
 6000 &  3.74 & -2.45 & -2.61 & 1.5 &  -0.181 &  3.32 &  3.14 & 4.40 &  1.95 &  0.76 &  1430 & ... & s?rs? & \\
   50 &  0.07 &  0.03 &  0.09 & 0.1 &	0.002 &  0.15 &  0.15 & 0.61 &  0.14 &  0.05 &   143 &&& \\
\hline
&&&&&\multicolumn{3}{c}{CS 29503-010} &&&&&& \\
 6050 &  3.66 & -1.69 & -1.70 & 1.6 &  -0.135 &  2.91 &  2.77 & 6.17 &  2.27 &  0.86 &  1400 &  0.12 & rs & \\
   50 &  0.08 &  0.02 &  0.06 & 0.1 &	0.003 &  0.15 &  0.15 & 0.87 &  0.17 &  0.05 &   146 &&& \\
\hline
&&&&&\multicolumn{3}{c}{CS 29512-073} &&&&&& \\
 5560 &  3.44 & -2.06 & -1.88 & 1.1 &  -0.195 &  2.78 &  2.58 & 7.35 &  2.93 &  0.86 &  1730 &  1.05 & s & \\
   50 &  0.07 &  0.02 &  0.04 & 0.1 &	0.006 &  0.12 &  0.12 & 0.85 &  0.18 &  0.02 &   160 &&& \\
\hline
&&&&&\multicolumn{3}{c}{CS 29526-110} &&&&&& \\
 6650 &  3.79 & -2.19 & -2.29 & 1.6 &  -0.152 &  2.85 &  2.70 & 6.62 &  1.95 &  0.85 &  1170 &  0.01 & rs & B \\
   50 &  0.07 &  0.04 &  0.10 & 0.1 &	0.002 &  0.12 &  0.12 & 0.76 &  0.12 &  0.02 &   108 &&& \\
\hline
&&&&&\multicolumn{3}{c}{CS 29528-028} &&&&&& \\
 7100 &  4.27 & -2.12 & -2.17 & 1.2 &  -0.155 &  3.81 &  3.65 & 2.75 &  1.10 &  0.81 &  1320 &  0.33 & rs & \\
   50 &  0.07 &  0.06 &  0.18 & 0.3 &	0.003 &  0.12 &  0.12 & 0.31 &  0.07 &  0.02 &   121 &&& \\
\hline
&&&&&\multicolumn{3}{c}{CS 31070-073} &&&&&& \\  
 6190 &  3.86 & -2.55 & -2.64 & 1.5 &  -0.181 &  3.43 &  3.25 & 3.99 &  1.74 &  0.80 &  1630 & -0.41 & r & \\
   50 &  0.07 &  0.03 &  0.08 & 0.1 &	0.001 &  0.12 &  0.12 & 0.46 &  0.11 &  0.02 &   150 &&& \\
\noalign{\smallskip}
\hline\hline
&\multicolumn{3}{c}{SAT subsample} &&&&&&&&&& \\
\hline
 $T_{\rm ex}$ (K) & $\log g$ & [Fe/H] & $\xi$ (km/s) & [Ba/Eu] & Class & Var. &&&&&&&& \\
\cline{1-7}
\noalign{\smallskip}
&\multicolumn{3}{c}{BS 16080-175} &&&&&&&&&& \\
 6240 $\pm$ 100 & 3.70 $\pm$ 0.1 & -1.86 $\pm$ 0.2 & 1.05 $\pm$ 0.1 &  0.5  & rs &&&&&&&&& \\
\cline{1-7}
\noalign{\smallskip}
&\multicolumn{3}{c}{BS 17436-058} &&&&&&&&&& \\
 5340 $\pm$ 100 & 2.20 $\pm$ 0.1 & -1.90 $\pm$ 0.2 & 2.70 $\pm$ 0.1 &  0.44 & rs &&&&&&&&& \\
\cline{1-7}
\noalign{\smallskip}
&\multicolumn{3}{c}{CS 22183-015} &&&&&&&&&& \\
 5470 $\pm$ 100 & 2.85 $\pm$ 0.1 & -2.85 $\pm$ 0.2 & 1.45 $\pm$ 0.1 &  0.52 & s?rs? &&&&&&&&& \\
\cline{1-7}
\noalign{\smallskip}
&\multicolumn{3}{c}{CS 22887-048} &&&&&&&&&&& \\
 6500 $\pm$ 100 & 3.35 $\pm$ 0.1 & -1.70 $\pm$ 0.2 & 2.05 $\pm$ 0.1 &  0.51 & s?rs? & B &&&&&&&& \\
\cline{1-7}
\noalign{\smallskip}
&\multicolumn{3}{c}{CS 22898-027} &&&&&&&&&& \\
 6240 $\pm$ 100 & 3.72 $\pm$ 0.1 & -2.30 $\pm$ 0.2 & 1.40 $\pm$ 0.1 &  0.35 & rs &&&&&&&&& \\
\cline{1-7}
\noalign{\smallskip}
&\multicolumn{3}{c}{CS 29502-092} &&&&&&&&&& \\
 4970 $\pm$ 100 & 1.70 $\pm$ 0.1 & -3.05 $\pm$ 0.2 & 1.85 $\pm$ 0.1 & -1.58 & r &&&&&&&&& \\
\cline{1-7}
\end{tabular}
}
\end{table*}
}

\subsection{The DMA subsample}

The adopted model atmospheres were computed with the latest version of
the ATLAS9 code, initially developed by \citet{kurucz92,kurucz93}, and
subsequently updated by \citet{castelli97} and \citet{castkur04}, with
enhanced $\alpha$-element abundances, but no C- or N- enhancement. The
current version uses an improved opacity distribution function, where
the solar abundances, TiO lines, and some atomic and molecular constants
were replaced. 

Before determining the photometric temperatures, an initial value for
interstellar reddening was estimated. Since there are no available
Hipparcos parallaxes for the stars in our sample, the first values for
the absolute magnitude ($M_{\rm V}$) of each star were taken from the
colour-magnitude diagrams (CMDs, $M_{\rm V}$) vs. B-V) given by
\citet{green87} and \citet{lcb98}. These diagrams usually provide two
$M_{\rm V}$ values for each B-V, according to the evolutionary state
of the star, so the first value to be used was
chosen arbitrarily. This absolute magnitude was then used to estimate
the first value for the distance; that, together with the Galactic
coordinates in Table \ref{fotomet}, were used to calculate the first
value for the visual extinction, $A_{\rm V}$, following
\citet{hakkila97}. Then, the photometric temperatures were calculated
using the colour-temperature calibrations of 
\citet{alon96a,alon99}, considering [Fe/H] = $-$1.5, $-$2, and $-$2.5. 
These authors use accurate stellar-diameter effective temperatures from the
literature to calculate temperatures for objects with observed multiple
pass-band colours, using the infrared flux-method (IRFM). \citet{alon96a,alon99}
use the Johnson system for $UBVRI$ and TCS (Telescopio Carlos S\'anchez) for
$JHKLMN$, so before determining the temperatures, the colours were transformed
according to \citet{bess79}, \citet{alon98}, and \citet{carp01}. The mean value
for the temperature derived from $B-V$, $V-R$, $V-I$, and colours derived from
2MASS magnitudes is the first value for the temperature used to create the
model atmosphere. For some stars, such as BS~16077-077, temperatures derived
from the $B-V$ and $V-K$ colours are very low compared to those from other
colours, so they were neglected. Bolometric corrections BC(V) were determined
according to \citet{alon95} or \citet{alon99}, depending on the more probable
evolutionary stage of the star, as estimated from the CMD. With the initial values of
temperature and extinction-corrected $(B-V)_o$, the first input for the mass was
taken from the Yonsei-Yale isochrones (Y$^2$), described by 
\citet{yi01}, \citet{kim02}, \citet{yi03} and \citet{dem04}. Then, the first 
estimate for $\log$ $g$ (and its error) were taken from the relations:

{\small
\begin{eqnarray}
\label{logtri}
\log\biggl({g_\ast\over g_\odot}\biggr) = & \log\biggl({M_\ast\over M_\odot}\biggr) +
4\log\biggl({T_{\rm eff\ast}\over T_{\rm eff\odot}}\biggr)+ 0.4V_\circ + \nonumber \\
 & + 0.4BC(V) + 2\log {1\over D} + 0.1
\end{eqnarray}

\begin{eqnarray}
\sigma_{\rm logg}= & \biggl[\biggl({\sigma_M\over
{M\ln(10)}}\biggr)^2+\biggl({4\sigma_{\rm Teff\ast}\over{T_{\rm eff\ast}\ln(10)}}\biggr)^2+\biggl({4\sigma_{\rm Teff\odot}\over
{T_{\rm eff\odot}\ln(10)}}\biggr)^2+ \nonumber \\
+ & \sigma_{\rm logg\odot}^{2}+(0.4\sigma_{V\circ})^2+(0.4\sigma_{BC})^2+\biggl({2\sigma_\pi\over{\pi\ln(10)}}\biggr)^2\biggr]^{0.5}
\end{eqnarray}
}

\noindent where $M_\ast$ is the stellar mass, $V_\circ$ is the
extinction-corrected magnitude, and $D$ is the distance. For the Sun, we have
adopted $T_{\rm eff\odot}$ = 5781 K \citep{bess98}; $\log g_\odot$ = 4.44;
$M_{\rm {bol\odot}}$ = 4.75 \citep{cram99}.

With the first estimates for $T_{\rm eff}$ and $\log g$, the \ion{Fe}{i}
and \ion{Fe}{ii} abundances were calculated using the code WIDTH9
\citep{cast04}, which derives abundances from the equivalent widths ({\it EW}) of atomic lines.
Table \ref{DMAiron} shows the \ion{Fe}{i} and \ion{Fe}{ii} lines, with
the respective atomic constants and $EW$ for each star of the DMA subsample. As
indicated in Table \ref{DMAiron}, the main source of oscillator
strengths and excitation potential for \ion{Fe}{i} lines was the library
of the National Institute of Standards \& Technology (NIST)
\citep{mart88,mart02}; for \ion{Fe}{ii}, values by \citet{fw06} were
adopted. A significant number of \ion{Fe}{i} lines were used, ranging
from 12 to 124. For two stars, less than 20 \ion{Fe}{i} lines were
available. Concerning \ion{Fe}{ii}, most stars had more than 10 lines
available. After running WIDTH9, we have the first estimates for the
\ion{Fe}{i} and \ion{Fe}{ii} abundances. Iron lines whose
abundances were beyond 1$\sigma$ from the overall mean value
were discarded for the next iteration. 

An isochrone for a given age changes if the metallicity changes, so we
have to look for new $M_{\rm V}$ and mass estimates at the isochrone of
metallicity closest to that obtained from the WIDTH9 analysis; the
process must then be restarted with this new $M_{\rm V}$, so that new
photometric temperatures can be derived. In this first iteration, it is
not unusual that the abundance of \ion{Fe}{i} is very different from
that of \ion{Fe}{ii}. One possible cause for this discrepancy is the
inadequacy of the initial choice of $M_{\rm V}$. This means that the
value corresponding to the (B-V) with opposite evolutionary state to the
first choice is the more suitable $M_{\rm V}$ value to be used.
Indeed, if [FeII/H] is higher than [FeI/H], $\log g$
must decrease in order to decrease [FeII/H]. In such a case,
the more suitable $M_{\rm V}$ is lower than the first
choice. Then, the process must be restarted using the new $M_{\rm V}$
and new parameters derived. When the difference between
$\log\epsilon$(FeI) and $\log\epsilon$(FeII) obtained with WIDTH9 is
below 0.2 dex, and close to those used in the previous iteration, fine
tuning adjustments are made. WIDTH9 provides fits for abundance {\it
vs}. excitation potencial ($\chi_{ex}$) and vs. {\it EW}, and analysing
their slopes for each iteration it is possible to seek the best set of
parameters. The temperatures were changed in the atmosphere model
depending on whether the $\chi$ slope was positive or negative. When
there is no trend between the abundances, $\chi_{ex}$, and {\it EW} we
have obtained our desired excitation temperature ($T_{\rm ex}$), and the
microturbulent velocity ($\xi$). 

Summarizing, the entire process is iterated until convergence is
obtained to a consistent set of parameters. The uncertainties on $T_{\rm
ex}$ and $\xi$ were estimated directly from this iterative process,
taking into account the slopes of the best fits. We had many iron lines,
hence the random uncertainties on \ion{Fe}{i} and
\ion{Fe}{ii} abundances were taken to be the standard error of the mean.
The various photometric temperatures derived with the final \ion{Fe}{i} and
\ion{Fe}{ii} abundances are shown in Table \ref{temps}, and all final parameters
are listed in Table \ref{paramDMA-SAT}. Photometric temperatures vary
considerably -- by up to a few hundred K -- for different colour indices, which
suggests a need for better photometry and/or transformations. It is worth noting
that the \citeauthor{alon99} effective temperatures were fairly accurate
for [Fe/H] $\gtrsim$ $-$2.5 to $-2.0$, but become non-physical for metallicities
smaller than $-3.0$, due to the small number of stars with [Fe/H] $\lesssim$
$-2.5$ available in the literature at the time the calibration was undertaken.
For this reason, we adopted $T_{\rm ex}$ as the temperatures for our sample of
stars in the DMA analysis.

Other stellar parameters and their uncertainties shown in Table
\ref{paramDMA-SAT} are derived through application of the equations below. 
Uncertainties on BC(V) are estimated by computing how the uncertainties
on $T_{ex}$ modify their values. From Eqs. \ref{magab} to \ref{raio},
one obtains the absolute magnitude $M_{\rm V}$, the bolometric magnitude
$M_{\rm bol}$, luminosity (L$_\ast$/L$_\odot$), and radius
($R_\ast$/R$_\odot$), providing as inputs
for the distance D(pc), $A_{\rm V}$, V, and $\log$ $g$:

\begin{equation}
\label{magab}
M_v = V-5\log D+5-A_v
\end{equation}
\begin{equation}
\sigma_{Mv}=\biggl[({\sigma_V)^2+\biggl({5\sigma_D\over D\ln(10)}\biggr)^2+\sigma_{Av}^2}\biggr]^{0.5}
\end{equation}
\begin{equation}
\label{magbol}
M_{bol\ast} = M_v+BC(V)
\end{equation}
\begin{equation}
\sigma_{Mbol\ast}=\biggl({\sigma_{Mv}^2+\sigma_{BC}^2}\biggr)^{0.5}
\end{equation}
\begin{equation}
\label{lumin}
L_\ast = 10^{-0.4(M_{bol\ast}-M_{bol\odot})}L_\odot
\end{equation}
\begin{equation}
\sigma_L=L0.4\ln(10)\biggl({\sigma_{Mbol\ast}^2+\sigma_{Mbol\odot}^2}\biggr)^{0.5}
\end{equation}
\begin{equation}
\label{raio}
R_\ast = \biggl({L_\ast T_{ef\odot}^4\over L_\odot T_{ef\ast}^4}\biggr)^{0.5}R_\odot
\end{equation}
\begin{equation}
\sigma_R=R\biggl[{\biggl({\sigma_L\over 2L}\biggr)^2+\biggl({2\sigma_{Teff\odot}\over T_{eff\odot}}\biggr)^2+\biggl({2\sigma_{Teff\ast}\over T_{eff\ast}}\biggr)^2}\biggr]^{0.5}.
\end{equation}

\begin{table}
\caption{Atomic lines for the SAT subsample with atomic constants and 
EW measurements (first four lines only; the full table is only available on CDS.)
The first four columns give 
wavelength, lower-level excitation potential, oscillator strength, and the source of the atomic data,
respectively. The remaining 11 columns 
give the equivalent width in m\AA\ of each feature measured in these stars. 
} 
\label{linesSAT}
\centering
\setlength\tabcolsep{3pt}
\begin{tabular}{llrrrrrrr}
\noalign{\smallskip}
\hline \hline
ion & $\lambda$ ($\rm \AA$) & $\chi_{ex}$ (eV) & $\log gf$ & Ref. & 1 & 2 & 3 & 4 \\
\hline
\noalign{\smallskip}
Ti I & 3729.81 & 0.00 & -0.30  & 7 & ... & ...    &  ...    &  ...    \\   
Ti I & 3741.07 & 0.02 & -0.16  & 7 & ... & 41.10  &  ...    &  ...    \\   
Ti I & 3924.53 & 0.02 & -0.88  & 7 & ... & ...    &  ...    &  74.69  \\   
Ti I & 3958.21 & 0.05 & -0.12  & 7 & ... & 65.77  &  ...    & 102.90  \\   
\noalign{\smallskip}
\hline
\end{tabular}

\tablefoot{Stars in columns 6 to 16: (1) BS~16080-175; (2) BS~16090-048; (3) BS~17436-058; (4) BS~17451-031;
(5) CS~22171-009; (6) CS~22174-007; (7) CS~22183-015; (8) CS~22887-048; (9) CS~22898-027; (10) CS~29502-092;
(11) HD 140283.
}

\tablebib{
(1) \citet{FMW88}; (2) \citet{kurucz75}; (3) NIST; (4) \citet{ryan95}; (5) \citet{T89}; 
(6) Kurucz (1989, private communication); (7) Experimental sources.
} 

\end{table}

\subsection{The SAT subsample}

The SAT analysis is fully described in \citet{tsang05}. For
completeness, we present here a short description of his procedure.
The grid of model atmospheres was taken from \citet{bell76} and Bell
(1983 - priv.comm.). The former work tabulates model atmospheres for
(early-G to early K-type) giants in the parameter space 3750 K $\leq$
$T_{\rm eff}$ $\leq$ 6000 K, 0.75 $\leq$ $\log g$ $\leq$ 3.0 and $-$3.0
$\leq$ [Fe/H] $\leq$ 0.0. The latter supplements this parameter space to
include dwarfs, main-sequence turnoff stars and subgiants; i.e., it extends
the upper limits of effective temperature and surface gravity to 6500 K
and 5.0 , respectively. Both works calculate the grid models by
scaling the solar abundance ratios of all naturally occurring elements
by each model's desired metallicitty.

For the SAT subsample, an earlier version of the code WIDTH9 was used,
WIDTH6 \citep{kurucz79}. Table \ref{linesSAT} shows the lines, as well
as the atomic constants and the {\it EW} used as input for the code. As
for the DMA subsample, a large number of \ion{Fe}{i} were used, ranging
from 44 to 105. For \ion{Fe}{ii}, the number of lines ranges from 7 to
17. SAT used two models for HD 140283. In the first, model A, the
temperature, gravity, microturbulence, and iron abundance were all
derived from the spectral analysis (requiring ionisation balance). This
resulted in a rather low gravity (see Table \ref{compSAT_DMA}), at odds
with the value obtained from Hipparcos parallax measurements.
Consequently, SAT calculated a second model, model B, in which the
temperature was determined from IRFM-calibrated colours (using Alonso et
al. 1996b), and the gravity was fixed at the parallax value. This
created a disparity in the abundances inferred from neutral and ionised
Fe lines, suggestive of overionisation. Model B is preferred, but
serves as a reminder that for stars without trigonometric parallaxes the
results could have greater uncertainties. In order to compare the SAT
and DMA procedures, the latter determined atmospheric parameters and
abundances of Ti II and C for the star HD~140283. Table
\ref{compSAT_DMA} shows that both procedures are in good agreement,
although the [C/Fe] abundance varies considerably between SAT (Model B) and the
DMA analysis, despite the similarity between the adopted parameters. The
initial model atmosphere for the WIDTH6 calculation was adopted by
estimating the metallicity and surface gravity of each target and
calculating its effective temperature from the observed colours. SAT
estimated the line-of-sight reddening E(B-V) of the targets' observed
colours using the dust maps of \citet{schlegel98}, as shown in Table
\ref{fotomet}. Then, dereddened colours were used in deriving the initial
effective temperatures. These were calculated using the semi-empirical
calibrations of \citet{alon96a} for dwarfs and \citet{alon99} for giants.
The initial surface gravity of the candidate CEMP stars was estimated
based on the approximate evolutionary stage of each target. For stars
with colours near the main-sequence turnoff (BS~16080-175,
BS~16090-048, BS~17451-031, CS~22171-009, CS~22887-048, and
CS~22898-027), SAT initially adopted $\log g$ = 4.0, and for objects with
colours on the giant branch (including BS~17436-058, CS~22174-007,
CS~22183-015, and CS~29502-092), $\log g$ = 2.5. An incorrect initial
gravity implied longer computation times for the abundances of Fe I and
Fe II to converge, but since the final adopted $\log g$ was set using
the Fe ionisation balance, it was not affected by the starting value.
The metallicity of the model has the smallest effect on the derived
abundances among these three atmospheric parameters, which was
illustrated by a posterior error analysis. Therefore, [Fe/H] = $-$1.5 for
all the initial models were adopted, which is the mean Population II
metallicity \citep{laird88}. 
All photometric temperatures and final parameters adopted for the SAT
subsample are shown in Tables \ref{temps} and \ref{paramDMA-SAT}.

\begin{table*}
\caption{Atmospheric parameters and abundances for HD 140283 obtained by SAT 
and DMA. Model (A) results are shown in column 2, model (B) results are 
shown in column 3, and the DMA results are shown in column 4. Column 5 shows the range of 
results from the references in column 6.}
\label{compSAT_DMA}
\centering
\begin{tabular}{cccccc}
\noalign{\smallskip}
\hline \hline
                 & SAT(A)           & SAT(B)          & DMA              & Lit.           & Ref. \\
\hline
\noalign{\smallskip}
$T_{\rm ex}$ (K) &  5680 $\pm$ 100  &  5690 $\pm$ 100  &  5690 $\pm$ 50   &  5550 to 5843  & 1 \\
$\log g$ (cgs)   &  3.10 $\pm$ 0.10 &  3.69 $\pm$ 0.10 &  3.63 $\pm$ 0.12 &  3.20 to 3.83  & 1 \\
$\xi$ (km/s)     &  1.5  $\pm$ 0.1  &  1.0  $\pm$ 0.1  &  1.5  $\pm$ 0.1  &  0.75 to 1.88  & 1 \\
{[FeI/H]}        & -2.65 $\pm$ 0.20 & -2.62 $\pm$ 0.20 & -2.66 $\pm$ 0.10  & -2.70 to -2.21 & 1 \\
{[FeII/H]}       & -2.65 $\pm$ 0.20 & -2.39 $\pm$ 0.20 & -2.44 $\pm$ 0.10  &  &  \\
{[C/Fe]}         &  0.59 $\pm$ 0.27 &  0.13 $\pm$ 0.10 &  0.71 $\pm$ 0.10  & 0.22 to 0.60 & 2,3,4,5 \\
{[TiII/Fe]}      &  0.11 $\pm$ 0.02 &  0.13 $\pm$ 0.10 &  0.23 $\pm$ 0.10  & 0.23 to 0.27 & 3,4 \\
 \noalign{\smallskip}
\hline
\end{tabular}

\tablefoot{The atmospheric parameters and abundances of model SAT(A) were calculated 
concurrently, while the IRFM effective temperature \citep{alon96b}
and parallax surface gravity were used in model SAT(B).
}

\tablebib{
(1) \citet{tsang05}; (2)\citet{tomkin92}; (3)\citet{norris97}; (4)\citet{aoki02e}; 
(5)\citet{akerman04}.
} 

\end{table*}

\section{Abundances and uncertainties}\label{abundt}

\subsection{Effectiveness of the selection method}

The stellar parameters and [Fe/H] values derived in Section~2 enable us
to examine the effectiveness of the selection criteria used for the CEMP
stars. The majority of these CEMP candidates are confirmed as metal-poor
stars. The five exceptions are as follows. BS~17451-031 has two
disparate $B-V$ colours, provided in Table \ref{fotomet}. The excitation
temperature (5800~K) is consistent with the more recent, bluer colour.
We infer that because of the older and presumably erroneous colour, viz.
B-V = 0.711, this Sun-like dwarf was originally mistaken as a cool giant
and, being weak lined compared to genuine giants, was misclassified as
metal-poor but, as with other Sun-like dwarfs, having a prominent
G-band. BS~16086-022 and CS~22169-002 were found to have extremely broad
H$_\gamma$, H$_\epsilon$ and $\lambda$4324 features, which reveal them
to be white dwarfs. The only narrow lines in these spectral regions are
the \ion{Ca}{ii} H and K lines, which are likely of interstellar origin.
For this reason, these two white dwarfs were misclassified as metal-poor
candidates in the HK-survey objective-prism selection. The spectra also
exhibit a broad, shallow depression of flux around 4310 $\rm \AA$. The
broad 4310 $\rm \AA$ and 4324 $\rm \AA$ depressions coincide with the CH
G-band, and suggest these stars to be C-rich (DQ) white dwarfs.
BS~16542-052 was listed amongst the emission line candidates of
\citet{beers94}. Our spectrum showed it to be strong lined, with a
preliminary analysis indicating [Fe/H] $\sim$ $-$0.8. This star is
unlikely to be a CEMP star, and is more likely a mildly metal-poor star
with a normal G-band. The \ion{Ca}{ii} H and K lines did not appear in emission at the
time of our observation. We chose not to analyse it further. Our
spectrum of BS~16088-104 showed this star to be quite strong lined, with
a preliminary analysis (conducted by A. Hosford) indicating [Fe/H]
$\sim$ $-0.3$. As it was almost certainly misclassified as a CEMP star;
we did not analyse it further.

We have found five misclassified stars in our echelle observations of 26
CEMP candidates, corresponding to a false-positive rate of 19\% in the
original sample selection. We do not consider these stars further.

%

\subsection{Other elemental abundances}\label{othabund} 

For the elements other than Fe, abundances were derived through 1D LTE
spectrum synthesis, using the code SYNTHE created by R. Kurucz, or in
the case of the SAT analysis, the synthesis code ATLAS by
\citet{cottrell78}. Abundances of carbon were derived using molecular
synthesis of the G-band (CH A$\sp 3$$\Delta$ - X$\sp 3$$\pi$) at
$\lambda\lambda$4290-4315\AA. Atomic lines were used to calculate the
abundances of Na, Mg, Sc, Ti, Cr, Cu, Zn, Sr, Y, Zr, Ba, La, Ce, Nd, Sm,
Eu, Gd, Dy by DMA, and Ti, Zn, Sr, Y, Zr, Ba, La, Eu, Pb by SAT. Table
\ref{linesSAT} shows the lines used for the SAT subsample; Table
\ref{abunDMA} lists the abundances line by line, along with the respective
atomic constants for the DMA subsample. 

The oscillator strengths for elements other than Fe, and their
respective sources, are shown in Table \ref{abunDMA}. For the $\alpha$-
and iron-peak elements, most values are from NIST; a few data are from
VALD \citep{pisk95,ryabch97, kupka99,kupka00}. Laboratory values, where
available, were given preference over theoretical values. For the lines
of \ion{Cu}{i}, \ion{Eu}{ii}, \ion{La}{ii}, and \ion{Ba}{ii} the
hyperfine structure ($hfs$) was taken into account, employing a code
made available by A. McWilliam, following the calculations described by
\citet{proc00}. The $hfs$ constants were taken from \citet{rut78} for
\ion{Ba}{ii}, \citet{law01a} for \ion{La}{ii}, and \citet{law01b} for
\ion{Eu}{ii}. The final $hfs$ components were determined by using the solar
isotopic mix by \citet{lod03} and the total $\log gf$ values from laboratory
measurements. For copper, the $hfs$ from \citet{bi76} was used, with isotopic
fractions of 0.69 for $^{63}$Cu and 0.31 for $^{65}$Cu. In this case, small
corrections were applied such that the total $\log gf$ equals the $gf$ value adopted
in this work. The lines for which $hfs$ were used were tested by creating a
sinthetic solar spectrum to compare to the solar observed spectrum by
\citet{kurucz84}.

Tables \ref{medDMA} and
\ref{medSAT} shows the average abundance for the elements in each star, for the DMA
and remaining SAT subsamples, respectively. For most elements, the solar
abundances used were extracted from \citet{gs98}; full references are
indicated in Table \ref{medDMA}. The usual notations $\log\epsilon$(A) =
$\log(N_A/N_H)$+12 and [A/B] = $\log(N_A/N_B) _\ast$-$\log(N_A/N_B)
_\odot$ were adopted.  

Uncertainties in the abundances were evaluated by consideration of the
degree to which a variation of 1$\sigma$ in the atmospheric parameters
and S/N affected the output value of the synthesis routine. For details
of the error calculations, see Appendix \ref{apenderr}.

Although the errors in the atmospheric parameters are usually the
dominant source of the total error on the abundance estimates,
observational sources can be more significant at a signal to noise ratio around
40. A change of 50~K in $T_{\rm eff}$ leads to a change of typically
0.03~dex in [X/Fe], with the largest change being 0.10~dex for some
elements. Changes of 0.10~dex in $\log g$ result in alterations of
[X/Fe] of up to 0.04~dex. Changes of 0.10 km/s in $\xi$ lead to [X/Fe]
changes up to 0.03~dex. For the present sample, the major contribution
to the uncertainties on [X/Fe] is the S/N ratio achieved in the spectra.
For example, for BS~16077-077 (the star in our sample with the highest
S/N) the contribution of error due to S/N ranges typically from 0.07~dex
to 0.15~dex, reaching, for some elements with very weak lines, as high
as 0.40~dex. For stars with S/N around 30, the contribution to the
uncertainties is even higher, ranging typically from 0.10~dex to
0.30~dex, reaching 0.50~dex for some elements. Uncertainties are listed
in Table
\ref{medDMA}.

\begin{table}
\caption{Lines, equivalent widths, and abundances for the DMA subsample.
See below for the $\log gf$ sources indicated in column 6.
The full table is only available at CDS.} 
\label{abunDMA} 
\centering
\setlength\tabcolsep{2pt}
\begin{tabular}{clrrrrrrr}
\noalign{\smallskip}
\hline \hline
Star & el & $\lambda$ & $\chi_{ex}$ & $\log gf$ & Ref. & EW & $\log\epsilon$ & [X/Fe] \\
\hline
\noalign{\smallskip}
BS 15621-047 &  C    & G-band   &  ...   &   ...   & ...& ... &  7.23 &  1.06 \\
BS 15621-047 & Na I  & 4982.808 &  2.100 &  -1.913 & 1  &  26 &  5.33 &  1.35 \\
BS 15621-047 & Na I  & 4982.813 &  2.100 &  -0.961 & 1  & ... &  ...  &  ...  \\
BS 15621-047 & Na I  & 5682.650 &  2.100 &  -0.700 & 1  & ... &  ...  &  ...  \\
BS 15621-047 & Na I  & 5688.193 &  2.104 &  -1.390 & 1  &  60 &  5.33 &  1.35 \\
\noalign{\smallskip}
\hline
\end{tabular}

\tablebib{
1. NIST; 2. \citet{fuhrmann95};
3. \citet{mart88}; 4. \citet{Bi75}; 
5. \citet{BG80}; 6. \citet{g94};
7. \citet{H82}; 8. \citet{HL83};
9. \citet{B81}; 10. \citet{T89};
11. \citet{S96}; 12. \citet{M98};
13. \citet{law01a}; 14. \citet{PQ00};
15. \citet{Hartog03}; 16. \citet{MW77}; 
17. \citet{B89}; 18. \citet{law01b}; 
19. \citet{C62}; 20. VALD;
21. \citet{B88}; 
22. average of \citet{K92} and \citet{BL93}. 
}
\end{table}

\onltab{8}{
\begin{table*}
{\scriptsize
\caption{Average of abundances for the DMA subsample of CEMP stars, and solar abundances used as reference, as well as 
their sources. Numbers in parenthesis are the errors in the last decimals.}
\label{medDMA}
\setlength\tabcolsep{3pt}
\begin{tabular}{rrrrrrrrrrrrrrrrrrrrrrrrr}
\hline
\noalign{\smallskip}
&& \multicolumn{2}{c}{BS 15621-047} && \multicolumn{2}{c}{BS 16033-081} && \multicolumn{2}{c}{BS 16077-077} && 
\multicolumn{2}{c}{BS 16082-129} && \multicolumn{2}{c}{BS 16543-097} && \multicolumn{2}{c}{BS 16929-005} \\ 
\cline{3-4} \cline{6-7} \cline{9-10} \cline{12-13} \cline{15-16} \cline{18-19} \\
el && $\log\epsilon$(X) & [X/Fe] && $\log\epsilon$(X) & [X/Fe] && $\log\epsilon$(X) & [X/Fe]
   && $\log\epsilon$(X) & [X/Fe] && $\log\epsilon$(X) & [X/Fe] && $\log\epsilon$(X) & [X/Fe] \\
\noalign{\smallskip}
\hline
\noalign{\smallskip}
 C    && 7.23 & 1.06(31) &&  7.22 &  0.77(36) &&  8.77 &  2.25(18) &&  7.13 &  1.20(30) &&  7.31 &  1.15(33) &&  7.00 & 1.50(23) &&&\\
Na    && 5.33 & 1.35(36) &&  4.83 &  0.57(29) &&  ...  &  ...	   &&  ...  &  ...	&&  ...  &  ...      &&  3.83 & 0.52(29) &&&\\
Mg    && 6.08 & 0.85(25) &&  5.58 &  0.07(32) &&  6.08 &  0.50(19) &&  ...  &  ...	&&  5.78 &  0.56(20) &&  5.28 & 0.72(20) &&&\\
Sc    && 1.37 & 0.55(37) &&  1.49 &  0.39(29) &&  1.67 &  0.50(16) &&  ...  &  ...	&&  ...  &  ...      &&  ...  & ...	 &&&\\
Ti    && 3.02 & 0.35(26) &&  3.41 &  0.46(22) &&  3.76 &  0.74(14) &&  3.02 &  0.59(26) &&  3.12 &  0.46(18) &&  2.72 & 0.72(22) &&&\\
Cr    && 3.67 & 0.35(14) &&  4.12 &  0.52(21) &&  4.07 &  0.40(13) &&  3.67 &  0.59(20) &&  3.17 & -0.14(17) &&  3.17 & 0.52(18) &&&\\
Cu    && ...  & ...	 &&  ...  &  ...      &&  ...  &  ...	   &&  ...  &  ...	&&  ...  &  ...      &&  ...  & ...	 &&&\\
Zn    && 3.86 & 1.61(48) &&  2.84 &  0.31(58) &&  3.18 &  0.58(30) &&  2.60 &  0.59(72) &&  2.60 &  0.36(48) &&  ...  & ...	 &&&\\
Sr I  && 1.87 & 1.25(37) &&  0.97 &  0.07(29) &&  ...  &  ...	   &&  ...  &  ...	&&  1.47 &  0.86(22) &&  ...  & ...	 &&&\\
Sr II && 0.86 & 0.24(20) &&  0.77 & -0.13(24) &&  0.91 & -0.06(16) && -0.03 & -0.41(30) &&  0.79 &  0.18(20) &&  0.47 & 0.52(23) &&&\\
 Y    && 0.40 & 0.51(15) && -0.02 & -0.19(19) &&  0.12 & -0.12(13) && -0.26 &  0.09(26) &&  0.36 &  0.48(15) &&  0.24 & 1.02(18) &&&\\
Zr    && 1.41 & 1.16(69) &&  0.60 &  0.07(19) &&  0.58 & -0.02(30) &&  0.60 &  0.59(70) &&  1.10 &  0.86(38) &&  1.10 & 1.52(50) &&&\\
Ba    && 0.13 & 0.35(21) && -0.87 & -0.93(18) &&  0.88 &  0.75(13) && -0.87 & -0.41(21) &&  0.13 &  0.36(16) && -0.62 & 0.27(18) &&&\\
La    &&-0.72 & 0.50(34) && -0.83 &  0.11(35) && -0.29 &  0.58(21) &&  ...  &  ...	&& -0.59 &  0.64(27) && -0.53 & 1.36(38) &&&\\
Ce    && 0.62 & 1.27(37) && -0.15 &  0.22(29) &&  0.30 &  0.60(16) && -0.22 &  0.67(37) && -0.10 &  0.56(22) &&  ...  & ...	 &&&\\
Nd    && 0.32 & 1.22(16) && -0.31 &  0.31(15) &&  0.05 &  0.60(12) && -0.50 &  0.64(16) &&  0.20 &  1.11(14) &&  0.00 & 1.57(15) &&&\\
Sm    && 0.19 & 1.53(30) &&  ...  &  ...      &&  ...  &  ...	   &&  ...  &  ...	&&  ...  &  ...      &&  ...  & ...	 &&&\\
Eu    &&-0.99 & 0.84(28) && -1.17 &  0.38(25) && -1.49 & -0.01(20) &&  ...  &  ...	&& -0.99 &  0.85(22) && -1.49 & 1.01(25) &&&\\
Gd    &&-0.06 & 1.17(29) && -0.38 &  0.57(31) && -0.88 &  0.00(21) &&  ...  &  ...	&& -0.88 &  0.36(28) &&  ...  & ...	 &&&\\
Dy    && 0.25 & 1.40(36) && -0.36 &  0.51(29) &&  ...  &  ...	   &&  ...  &  ...	&&  ...  &  ...      &&  ...  & ...	 &&&\\
\hline\hline
\noalign{\smallskip}
&& \multicolumn{2}{c}{CS 22949-008} && \multicolumn{2}{c}{CS 29503-010} && \multicolumn{2}{c}{CS 29512-073} && 
\multicolumn{2}{c}{CS 29526-110} && \multicolumn{2}{c}{CS 29528-028} && \multicolumn{2}{c}{CS 31070-073} && \multicolumn{2}{c}{SUN}\\ 
\cline{3-4} \cline{6-7} \cline{9-10} \cline{12-13} \cline{15-16} \cline{18-19} \cline{21-22} \\
el && $\log\epsilon$(X) & [X/Fe] && $\log\epsilon$(X) & [X/Fe] && $\log\epsilon$(X) & [X/Fe]
   && $\log\epsilon$(X) & [X/Fe] && $\log\epsilon$(X) & [X/Fe] && $\log\epsilon$(X) & [X/Fe] && $\log\epsilon$(X) & Ref. \\
\noalign{\smallskip}
\hline
\noalign{\smallskip}
 C    &&  7.91 &  2.00(25) &&  8.47 &  1.65(20) &&  8.04 &  1.40(25) &&  8.61 &  2.38(20) &&  9.11 &  2.76(30) &&  7.22 &  1.34(22) && 8.52(6) & 1 \\
Na    &&  ...  &  ...      &&  5.44 &  0.81(22) &&  ...  &  ...      &&  ...  &  ...  	  &&  ...  &  ...      &&  ...  &  ...      && 6.33(3) & 1 \\
Mg    &&  ...  &  ...      &&  6.08 &  0.20(20) &&  6.08 &  0.38(20) &&  ...  &  ...  	  &&  ...  &  ...      &&  5.58 &  0.64(20) && 7.58(5) & 1 \\
Sc    &&  ...  &  ...      &&  2.27 &  0.80(22) &&  1.67 &  0.38(37) &&  ...  &  ...  	  &&  ...  &  ...      &&  ...  &  ...      && 3.17(0) & 1 \\
Ti    &&  3.02 &  0.61(22) &&  4.05 &  0.73(18) &&  3.52 &  0.38(26) &&  ...  &  ...  	  &&  ...  &  ...      &&  ...  &  ...      && 5.02(6) & 1 \\
Cr    &&  3.17 &  0.11(18) &&  4.17 &  0.20(15) &&  ...  &  ...      &&  ...  &  ...  	  &&  ...  &  ...      &&  ...  &  ...      && 5.37(3) & 1 \\
Cu    &&  ...  &  ...      &&  2.61 &  0.10(22) &&  ...  &  ...      &&  ...  &  ...  	  &&  ...  &  ...      &&  ...  &  ...      && 4.21(4) & 1 \\
Zn    &&  ...  &  ...      &&  3.54 &  0.64(30) &&  3.10 &  0.38(41) &&  ...  &  ...  	  &&  ...  &  ...      &&  ...  &  ...      && 4.60(8) & 1 \\
Sr I  &&  ...  &  ...      &&  2.47 &  1.20(22) &&  1.97 &  0.88(37) &&  ...  &  ...  	  &&  2.97 &  2.17(29) &&  2.07 &  1.74(29) && 2.97(7) & 1 \\
Sr II &&  0.19 & -0.17(25) &&  2.32 &  1.05(18) &&  1.71 &  0.62(20) &&  0.97 &  0.29(19) &&  2.07 &  1.27(27) &&  0.29 & -0.04(21) && 2.97(7) & 1 \\
 Y    &&  0.88 &  1.25(18) &&  1.63 &  1.09(12) &&  0.96 &  0.60(16) &&  1.24 &  1.29(20) &&  2.06 &  1.99(26) &&  1.56 &  1.96(20) && 2.24(3) & 1 \\
Zr    &&  1.60 &  1.61(60) &&  2.16 &  1.26(35) &&  1.31 &  0.59(50) &&  2.10 &  1.79(50) &&  2.60 &  2.17(60) &&  1.92 &  1.96(60) && 2.60(2) & 1 \\
Ba    &&  0.50 &  0.98(18) &&  2.24 &  1.81(16) &&  1.67 &  1.42(21) &&  2.13 &  2.29(16) &&  2.45 &  2.49(18) &&  1.91 &  2.42(18) && 2.13(5) & 1 \\
La    &&  ...  &  ...      &&  1.59 &  2.16(20) &&  0.67 &  1.42(30) &&  1.41 &  2.57(30) &&  1.17 &  2.21(45) &&  1.09 &  2.60(24) && 1.13(3) & 2 \\
Ce    &&  0.58 &  1.49(29) &&  2.05 &  2.05(22) &&  1.45 &  1.63(37) &&  1.70 &  2.29(22) &&  2.00 &  2.47(29) &&  1.71 &  2.65(29) && 1.70(4) & 3 \\
Nd    &&  0.82 &  1.98(15) &&  2.06 &  2.31(14) &&  1.14 &  1.57(16) &&  1.35 &  2.19(14) &&  1.82 &  2.54(15) &&  1.33 &  2.52(15) && 1.45(1) & 4 \\
Sm    && -0.99 &  0.61(30) &&  1.65 &  2.34(15) &&  1.01 &  1.88(37) &&  ...  &  ...  	  &&  ...  &  ...      &&  0.71 &  2.34(30) && 1.01(6) & 5 \\
Eu    &&  ...  &  ...      &&  0.51 &  1.69(22) && -0.99 &  0.37(28) &&  0.51 &  2.28(22) &&  0.51 &  2.16(25) &&  0.71 &  2.83(25) && 0.52(1) & 6 \\
Gd    &&  ...  &  ...      &&  ...  &  ...      &&  0.62 &  1.38(39) &&  ...  &  ...  	  &&  ...  &  ...      &&  ...  &  ...      && 1.12(4) & 1 \\
Dy    &&  ...  &  ...      &&  0.55 &  1.05(22) &&  0.64 &  1.32(36) &&  ...  &  ...  	  &&  ...  &  ...      &&  ...  &  ...      && 1.20(6) & 7 \\
\end{tabular}
}
\tablebib{
(1) \citet{gs98}; (2) \citet{law01a}; (3) \citet{PQ00};
(4) \citet{Hartog03}; (5) \citet{B89}; (6) \citet{law01b};
(7) \citet{BL93}.
}
\end{table*}
}

\onltab{9}{
\begin{table*}
{\scriptsize
\caption{Average of abundances for the SAT subsample of CEMP stars.}
\label{medSAT}
\begin{tabular}{lrrrrrrrrrrrrrrrrrr}
\hline\hline
\noalign{\smallskip}
 && \multicolumn{2}{c}{BS 16080-175} && \multicolumn{2}{c}{BS 17436-058} && \multicolumn{2}{c}{CS 22183-015} &&
\multicolumn{2}{c}{CS 22887-048} && \multicolumn{2}{c}{CS 22898-027} \\
\noalign{\smallskip}
\cline{3-4} \cline{6-7} \cline{9-10} \cline{12-13} \cline{15-16} \\
el && $\log\epsilon$(X) & [X/Fe] && $\log\epsilon$(X) & [X/Fe] && $\log\epsilon$(X) & [X/Fe]
   && $\log\epsilon$(X) & [X/Fe] && $\log\epsilon$(X) & [X/Fe] \\
\noalign{\smallskip}
\hline
\noalign{\smallskip}
C    && -0.01   &   1.75 && -0.28 & 1.50 &&   -0.51 &   2.34 &&    0.14 &   1.84 &&  0.04 & 2.34 \\
N    &&$<$-1.06 &$<$0.80 && -0.28 & 1.25 &&   -1.06 &   1.79 &&$<$-0.41 &$<$1.29 && -1.06 & 1.24 \\
TiI  && -1.22   &   0.64 && -1.59 & 0.19 &&   -2.38 &   0.47 &&   ...   & ...    &&  ...  & ...  \\
TiII && -1.56   &   0.30 && -1.32 & 0.46 &&   -2.63 &   0.22 &&   -1.22 &   0.48 && -1.98 & 0.32 \\
Zn   && -1.66   &   0.20 && -1.77 & 0.01 &&$<$-2.51 &$<$0.34 &&$<$-0.93 &$<$0.77 && -1.76 & 0.54 \\
Sr   && -0.82   &   1.04 && -0.83 & 0.95 &&   -2.31 &   0.54 &&   -0.70 &   1.00 && -1.71 & 0.59 \\
Y    && -0.79   &   1.07 && -1.05 & 0.73 &&   -2.31 &   0.54 &&   -0.71 &   0.99 && -1.70 & 0.60 \\
Zr   && -0.57   &   1.29 && -0.85 & 0.93 &&   -2.11 &   0.74 &&   -0.47 &   1.23 && -1.02 & 1.28 \\
Ba   && -0.31   &   1.55 && -0.17 & 1.61 &&   -0.96 &   1.89 &&    0.30 &   2.00 && -0.04 & 2.26 \\
La   && -0.21   &   1.65 && -0.29 & 1.49 &&   -1.32 &   1.53 &&    0.03 &   1.73 && -0.11 & 2.19 \\
Eu   && -0.81   &   1.05 && -0.61 & 1.17 &&   -1.48 &   1.37 &&   -0.21 &   1.49 && -0.39 & 1.91 \\
Pb   &&  0.74   &   2.60 &&  0.48 & 2.26 &&    0.15 &   3.00 &&    1.70 &   3.40 &&  0.59 & 2.89 \\
\noalign{\smallskip}
\hline\hline
\noalign{\smallskip}
 && \multicolumn{2}{c}{CS 29502-092} && & && \multicolumn{2}{c}{BS 16090-048} && \multicolumn{2}{c}{CS 22171-009} && 
\multicolumn{2}{c}{CS 22174-007} &&\\
\noalign{\smallskip}
\cline{3-4} \cline{9-10} \cline{12-13} \cline{15-16} \\
el && $\log\epsilon$(X) & [X/Fe] && & && $\log\epsilon$(X) & [X/Fe] && $\log\epsilon$(X) & [X/Fe] && $\log\epsilon$(X) & [X/Fe] \\
\noalign{\smallskip}
\hline
\noalign{\smallskip}
C    &&   -1.86 &    1.18 &&&&& -0.70 & 0.69 &&$<$-1.53 & $<$1.03 &&   -2.46 &   0.25 \\
N    &&   -1.76 &    1.28 &&&&& -0.55 & 0.84 &&$<$-0.03 & $<$2.53 &&$<$-2.11 &$<$0.60 \\
TiI  &&   -2.98 &    0.06 &&&&& -1.14 & 0.25 && ...     &  ...    &&   -2.43 &   0.28 \\
TiII &&   -2.95 &    0.09 &&&&& -0.93 & 0.46 &&   -2.27 &    0.29 &&   -2.33 &   0.38 \\
Zn   &&   -2.86 &    0.18 &&&&& ...   & ...  &&$<$-2.02 & $<$0.54 &&   -2.56 &   0.15 \\
Sr   &&   -3.54 &   -0.50 &&&&& ...   & ...  &&   -2.60 &    0.04 &&   -2.93 &  -0.22 \\
Y    &&   -3.54 &   -0.50 &&&&& ...   & ...  &&$<$-2.29 & $<$0.27 &&   -3.02 &  -0.31 \\
Zr   &&$<$-3.41 &$<$-0.37 &&&&& ...   & ...  &&$<$-1.76 & $<$0.80 &&   -2.67 &   0.04 \\
Ba   &&   -4.41 &   -1.37 &&&&& ...   & ...  &&   -2.83 &   -0.27 &&   -2.75 &  -0.04 \\
La   &&$<$-3.10 & $<$0.06 &&&&& ...   & ...  &&$<$-1.58 & $<$0.98 &&   -2.75 &  -0.04 \\
Eu   &&$<$-2.83 & $<$0.21 &&&&& ...   & ...  &&$<$-0.73 & $<$1.83 &&   -2.41 &   0.30 \\
Pb   &&$<$-1.51 & $<$1.53 &&&&& ...   & ...  &&$<$-0.49 & $<$2.07 &&$<$-1.43 &$<$1.28 \\
\noalign{\smallskip}
\hline
\end{tabular}
}
\end{table*}
}

\onltab{10}{
\begin{table*}
\caption{Comparison between results from this work and from the literature.} 
\label{compar} 
\centering
\begin{tabular}{lrlllllrrrlc}
\noalign{\smallskip}
\hline \hline
Star & $T_{\rm eff}$(K) & $\log$g & [FeI/H] & [FeII/H] & $\xi$(km/s) & [C/Fe] & [Ba/Fe] & [Eu/Fe] & [Ba/Eu] & class & Ref. \\
\hline
\noalign{\smallskip}
BS 16082-129  & 4900 & 1.8  & -2.86 & ...   & 1.6  & 0.29 & -0.97 & ...  &  ...   &    & 1 \\
	      & 5080 & 1.98 & -2.65 & -2.59 & 1.30 & 1.20 & -0.87 & ...  &  ...   &    & DMA \\
\hline
\noalign{\smallskip}
BS 16929-005  & 5270 & 2.7  & -3.09 & ...   & 1.3  & 0.92 & -0.59 & ...  &  ...   &    & 1 \\
              & 5250 & 2.8  & -3.2  & ...   & 2.3  & 1.08 & -0.48 & ...  &  ...   &    & 2 \\
	      & 5250 & 3.10 & -3.15 & -3.02 & 0.90 & 1.50 &  0.27 & 1.01 &  -0.74 & r  & DMA  \\
\hline
\noalign{\smallskip}
CS 29503-010  & 6500 & 4.5  & -1.2  & ...   & 1.9  & 1.07 &  1.50 & ...  &  ...   &    & 2 \\
              & 6050 & 3.66 & -1.69 & -1.70 & 1.6  & 1.65 &  1.81 & 1.69 &   0.12 & rs & DMA  \\
\hline
\noalign{\smallskip}
CS 29526-110  & 6500 & 3.2  & -2.38 & ...   & 1.6  & 2.20 &  2.11 & 1.73 &   0.38 & rs & 3, 8 \\
              & 6800 & 4.1  & -2.1  & ...   & 2.1  & 2.08 &  2.39 & ...  &  ...   &    & 4 \\
              & 6650 & 3.79 & -2.19 & -2.29 & 1.6  & 2.38 &  2.29 & 2.28 &   0.01 & rs & DMA  \\
\hline
\noalign{\smallskip}
CS 29528-028  & 6800 & 4.0  & -2.9  & ...   & 1.9  & 2.77 &  3.27 & ...  &  ...    &    & 2 \\
              & 7100 & 4.27 & -2.12 & -2.17 & 1.2  & 2.76 &  2.49 & 2.16 &      0.33 & rs & DMA  \\
\hline
\noalign{\smallskip}
CS 22183-015  & 5200 & 2.0  & -3.12 & ...   & 1.5  & ...  &  2.09 & 1.39 &   0.70 & s  & 5 \\
              & 5470 & 2.85 & -2.85 & ...   & 1.45 & 2.34 &  1.89 & 1.37 &   0.52 & rs & SAT  \\
\hline
\noalign{\smallskip}
CS 22898-027  & 6250 & 3.7  & -2.26 & ... & 1.5  & 2.2  & 2.23 & 1.88 & 0.35 & rs & 3,8 \\
	      & 6300 & 4.0  & -2.0  & ... & 2.0  & 1.95 & 2.27 & 1.94 & 0.33 & rs & 6 \\
              & 6000 & 3.6  & -2.36 & ... & 1.42 & 1.90 & 2.67 & 2.04 & 0.58 & s  & 9  \\
              & 6240 & 3.72 & -2.30 & ... & 1.40 & 2.34 & 2.26 & 1.91 & 0.35 & rs & SAT  \\
\hline
\noalign{\smallskip}
CS 29502-092  & 5000 & 2.1  & -2.76 & ...   & 1.8  & 1.0  & -0.82 &$>$0.4 &   -1.22 & no & 7\\
              & 4970 & 1.70 & -3.05 & ...   & 1.85 & 1.18 & -1.37 &$<$0.21&$>$-1.58 & no & SAT  \\
\hline
\noalign{\smallskip}
\end{tabular}

\tablebib{
1 - \citet{honda04}; 
2 - \citet{aoki07}; 
3 - \citet{aoki02d}; 
4 - \citet{aoki08}; \\
5 - \citet{johnson02}; 
6 - \citet{preston01}; 
7 - \citet{aoki02a};
8 - \citet{aoki02c};
9 - \citet{mcwill95}.
}
\end{table*}
}

%

\section{Discussion}\label{disc}

In order to study the distributions of elemental abundances for as many
known CEMP stars as possible, an extensive search for abundances
available in the literature has been performed. Elemental abundances for
C, N, O, Na, Mg, Al, Si, Ca, Sc, Ti, V, Cr, Mn, Co, Ni, Zn, Sr, Y, Zr,
Ba, La, Ce, Pr, Nd, Sm, Eu, Dy, and Pb (see Tables \ref{alpha} to
\ref{heavy2}) are shown in the figures, together with abundances for our
sample stars. The following criteria were applied to the literature
data: 
 
\begin{itemize}
   \item only data with resolving power R $>$ 30,000 were used
   \item if two works belong essentially to the same group of scientists, the
more recent value was adopted
   \item the average abundance was taken for the same star studied by different groups
   \item an upper limit was used only when there was no other value available
\end{itemize}

One difficulty in combining data from the literature is that we no
longer have a homogeneous abundance analysis, since different authors
employ different procedures which could involve different line lists,
atomic data for the lines used, adopted solar abundances, model
atmospheres, and spectrum synthesis codes. All of these could contribute
to an increase in the scatter in the patterns of observed abundances.

Eight stars from our observational program are in common with other
works from the literature meeting the criteria above, as shown in Table
\ref{compar}. However, among those, only four had an Eu abundance
determined elsewhere. For the others, Eu abundances were determined here
for the first time. We have chosen not to average our measurements with
others from the literature.
 
\subsection{C-rich stars: selection and classification} 
   
Carbon abundances for the 21 stars for which we have analysed C (Tables
\ref{medDMA} and \ref{medSAT}) are in the range +0.25 $\le$ [C/Fe] $\le$
+2.76. Of these, only three (BS~16033-081, BS~16090-048, and
CS~22174-007) have [C/Fe] $<$ +1.0, falling below the formal threshold
for classification as a CEMP star 
\citep[the usual definition for CEMP objects is {[C/Fe]} $>$ +1.0;][]{beers05}, 
although even two of these have [C/Fe] $>$ +0.6, i.e. four times the solar ratio. 
In addition, the star CS~22171-009 has an upper limit of [C/Fe] $<$ +1.03,
so it may not be C-rich. 

At face value, this suggests a further 14\% misclassification (false
positives) in the CEMP-star selection, or overall a $\sim$70\% success
rate in identifying genuine CEMP objects. However, according to
\citet{aoki07}, some stars with [C/Fe] $<$ +1.0 could still be CEMP
stars, depending on their evolutionary stage, since the surface carbon
abundance is expected to decrease after first dredge-up. Following their
criteria, CS~30322-023 was classified by them as CEMP star, although it
has [C/Fe] = +0.56. This is one of the stars from the literature
represented by open triangles in Fig. \ref{15615fg1}. Other stars from the
literature with low [C/Fe] that could still be considered CEMP stars are
shown in Fig. \ref{15615fg1}: CS~22885-096 with [C/Fe] = +0.60
\citep[cross,][]{norris01}, CS~30314-067 with [C/Fe] = +0.50
\citep[cross,][]{aoki02a}, BS~17435-532 with [C/Fe] = +0.68 
\citep[filled square,][]{roederer08}. Applying the \citeauthor{aoki07} criteria for 
stars of this work, BS~16033-081 with [C/Fe] = +0.77 also can be considered
C-rich. Consequently, 18 among 21 stars of our sample are C-rich.

Large over-abundances of C and N in CEMP stars are usually explained by
one of three mechanisms: mass transfer from a companion AGB star in a
binary system (environmental); self-enrichment of the star in these
elements (internal); or from the star having been born from a
previously-enriched cloud (congenital). Mechanisms to produce such large
amounts of C and N in population III stars have been studied by, e.g.,
\citet{fujimoto00} and \citet{siess02}, who claim that if mixing is
efficient during the He-flash in the core when the star reaches the tip
of the red-giant branch, C and N abundances can be significantly
enhanced. However, that mechanism cannot be the origin of internal C and
N enrichment in stars that have not reached this evolutionary stage, as
can be seen from $\log g$ in Table
\ref{alpha}. For example, if the evolutionary stage of the very C- and N-rich
star HE~1327-2326 is a dwarf or subgiant 
\citep[present evidence suggests a subgiant classification;][]{korn09}, 
this mechanism cannot be applied, and the source of these elements must
be extrinsic (environmental or congenital). Very massive stars (e.g.,
Wolf-Rayet stars) undergo significant mass loss during the red
supergiant stage, releasing C and N to the interstellar medium.
\citet{hirschi07} claim that the CNO abundances of HE~1327-2326 can be
reproduced by the winds of their rotating massive-star models.
\citet{umeda03} have simulated the supernova explosion of population III
progenitors from 20 to 130~M$_\odot$ and calculated the expected
nucleosynthesis yields. In their 25M$_\odot$ model, CNO are produced
during the pre-supernovae phase, in He shell burning, and although N is
underproduced in this stage, it could have already been produced through
the C-N cycle, and brought up to the surface during the first dredge-up
stage. In this model, only a small amount of the heavier elements such
as Mg, Ca, and the iron-peak elements are produced.
\citeauthor{umeda03} claim that the elemental abundances for this model are in
good agreement with the abundances of HE~0107-5240 by \citet{christ02,christ04},
characterized by very large [C/Fe] and [N/Fe] ratios, while the abundances of
elements heavier than Mg are normal.

Because of the relation between the r-process and massive star evolution 
\citep{truran81,ww95}, the expectation is that there might be evidence of 
element production from the r-process in the early Galaxy. In contrast, due to the
secondary nature of the s-process, the naive expectation is that evidence of this
source in metal-deficient stars should be scarce \citep{truran81,travaglio99}.
However, many CEMP stars have been found to exhibit clear s-process signatures
\citep{norris97}. Given that Ba and Eu are significant
representatives of the s- and r-processes in Solar System material,
CEMP-star classifications emerged to reflect their neutron-capture element compositions.
\citet{beers05} defined subclasses of CEMP stars as follows:
\medskip

CEMP-r: [Eu/Fe] $>$ +1.0

CEMP-s: [Ba/Fe] $>$ +1.0 and [Ba/Eu] $>$ +0.5

CEMP-r/s: 0 $<$ [Ba/Eu] $<$ +0.5

CEMP-no: [Ba/Fe] $<$ 0.

\medskip
\citet{aoki02b} obtained large over-abundances of C, N, Mg and Si from their 
analysis of CS~29498-043, and suggested a new class of CEMP stars --- CEMP-$\alpha$.

The [Ba/Eu] ratio for the stars from the literature was determined (see
Table \ref{alpha}). Then, following the \citeauthor{beers05} definition,
we found that 19 stars in the literature sample are CEMP-s, 2 are
CEMP-r, 15 are CEMP-r/s, and 15 might be CEMP-no, although among the
latter only one has had its Eu abundance determined, 7 have only an
upper limit, and 7 have no Eu abundance estimated; such stars are
represented by different symbols in Figs. \ref{15615fg1} -- \ref{15615fg7}, 
and their stellar classifications are indicated in Table \ref{alpha}. 
Five stars among the 109 stars cannot be classified yet, since they have no published 
Ba or Eu abundance.

\subsection{The $\alpha$- and iron-peak elements} 

\begin{figure*}
\centering
\includegraphics[width=16cm]{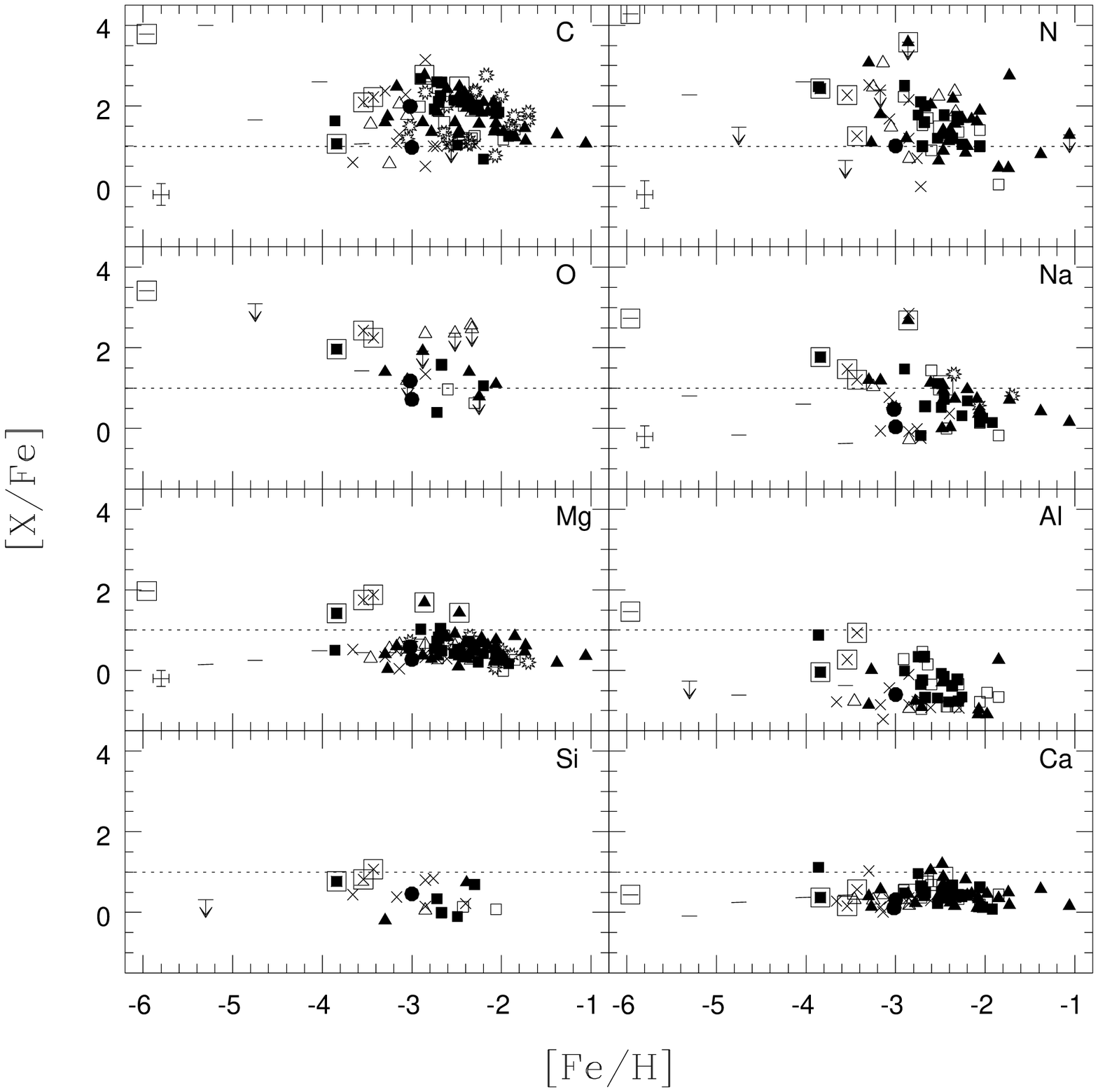}
\caption{[X/Fe] for CEMP stars of this work and from the literature. 
Symbols -- 
starred circles: this work;
filled squares: 0 $<$ [Ba/Eu] $\lesssim$ 0.5 (CEMP-r/s);
open squares: [Ba/Eu] $>$ 0.5 (CEMP-s);
crosses: [Ba/Fe] $<$ 0 (CEMP-no);
filled triangles: [Ba/Fe] $>$ +1 but there is no available Eu abundance (s or r/s?); 
open triangles: 0 $<$ [Ba/Fe] $<$ +1 but there is no available Eu abundance (r?);
filled circles: [Ba/Eu] $<$ 0 (r); dashes: upper limit for Ba and Eu; big open squares 
involving four crosses, two filled triangles and one dash indicate high over-abundance of Mg, 
[Mg/Fe] $>$ +1.4.
The dotted lines indicate where [X/Fe] = 1. See text for an explanations
concering stars with [C/Fe] $<$ +1.}
\label{15615fg1}
\end{figure*}

Figure \ref{15615fg1} shows that the runs of [X/Fe] vs. [Fe/H] for most of
the CEMP stars follow the canonical trend of metal-poor stars for O, Na,
Mg, Al, Si, and Ca (Tsangarides 2005 did not compute abundances for Na,
nor for V, Mn, Co, or Ni). The Mg abundance for the CEMP stars in Fig. \ref{15615fg1} 
remains around [Mg/Fe] = +0.5, according to the canonical
$\alpha$-element behaviour, except for six stars with [Fe/H] $<$
$-$2.40, for which very high Mg abundances are found, [Mg/Fe] $>$ +1.4
(see Table \ref{hiab}): CS~22949-037, CS~29498-043, CS~29528-028,
HE~1447+0102, HE~1327-2326, and HE~1012-1540. These six stars are
outlined in Fig. \ref{15615fg1} by large open squares, where they are seen
to also have [Na/Fe] $\gtrsim$ +1.2. In addition, most of the Mg-rich
stars also have high [Al/Fe] ratios (Fig. \ref{15615fg1}), so the
CEMP-$\alpha$ class would appear to be enriched in a wider set of
elements including Na, Mg and Al, implying that a ``CEMP-light''
classification might be more appropriate than CEMP-$\alpha$. In fact,
both of these suggested sub-classes are also CEMP-no stars, so the
nomenclature should be revisted. 

The stars CS~22942-019, CS~29497-034, SDSS~1707+58, CS~22958-042, shown
in Table \ref{hiab}, also have [Na/Fe] $\geq$ +1.4, although their
[Mg/Fe] ranges from +0.3 to +1.1. The star with a high Na abundance
represented by a filled triangle in Fig. \ref{15615fg1} without a big open
square surrounding it is SDSS 1707+58, for which \citet{aoki08}
determined [Na/Fe] = +2.71 and [Mg/Fe] = +1.13. One star of the present
sample, BS~15621-047, also has a high Na abundance, with [Na/Fe] = +1.35
and [Mg/Fe] = +0.85 (see Table \ref{medDMA}).

As seen in Table \ref{hiab}, some CEMP stars exhibit large
over-abundances of the lighter elements N, O, Na, Mg, and Si, in
particular among the lowest metallicity stars \citep{aoki07}. This Table
suggests that all CEMP stars with high [Mg/Fe] and [Na/Fe] are also
oxygen-rich, with [O/Fe] $\gtrsim$ +1.0. Unfortunately, some of these
stars have no oxygen abundance available in the literature to confirm
this point. It is worth noting that other stars in Table \ref{hiab},
CS~29497-030, CS~29528-041, CS~31080-095, and HE~0557-4840 are O-rich,
but are not Na- or Mg-rich. Three of the stars in Table \ref{hiab} with
high Mg abundances have also the highest Si abundances, with [Si/Fe]
ranging from +0.8 to +1.1. However, the last two stars, CS~30314-067 and
CS~29502-092, have [Si/Fe] in this range, but their Mg, Na and O
abundances are low. In contrast to O and Na, C and N seem to have no
correlation with the high over-abundance of Mg. Figure \ref{15615fg5}c
makes the pattern of [Mg/Fe] vs. [C/Fe] clearer, where one can see that
stars with the highest Mg abundance are spread over the range +1.0
$\lesssim$ [C/Fe] $\lesssim$ +3.8.

The [Ca/Fe] ratios for all of these stars are not as high as [O/Fe],
[Na/Fe], and [Mg/Fe], and the Ca abundance has no relationship with a
high Mg abundance. Figure \ref{15615fg1} shows that [Ca/Fe] vs. [Fe/H]
follows a flat pattern, and Fig. \ref{15615fg2} shows that the same
occurs for the abundances of elements lighter than Zn for different
classes of CEMP stars, with only a few exceptions. HE~1327-2326 has only
upper limits for V, Zn, Mn, and Co. For other elements, the CEMP star
abundances follow the general trend formed by other stars. Table
\ref{hiab} also indicates that there is no segregation among stars with high
over-abundance of the $\alpha$-elements according to the class of CEMP star,
since one can find several classes of CEMP stars with this characteristic.

Large over-abundances relative to iron for these intermediate-mass
elements cannot be explained by (environmental) mass transfer from a
companion AGB star. \citet{turatto98} and \citet{zampieri03} used the
fallback supposition to explain the observed low Fe abundance for SN
1997D and SN 1999br, respectively, and claim that similar SNe may be the
origin of the abundance pattern of stars with high abundances of Na, Mg,
or Si relative to iron-peak elements. These supernovae, characterized by
relatively small Fe ejection, are massive, but less energetic, and very
underluminous \citep{ww95}, the so-called ``faint supernovae''. In this
view, extremely metal-poor stars without enhancements of C and N
correspond to the abundance patterns from energetic supernovae
\citep{umeda03}. CEMP stars with large over-abundances of
intermediate-mass elements may have similarly formed, in a congenital
scenario, from gas enriched by SN experiencing partial fallback of
heavier material onto the collapsed remnant. In this case, Ca, Fe, as
well as other Fe-group elements such as Cr, Mn, Co, and Ni produced in
the deep interior, would fall back onto the nascent neutron star. In
contrast, lighter elements synthesized in the outer layers of the star
would be ejected and be excluded from any fallback. As a result, lighter
elements are significantly enhanced with respect to Fe-group elements.
\citet{aoki04} and \citet{cohen08} claim that the observed elemental abundance patterns of
CS~29498-043 and CS~22949-037, respectively, exhibit characteristics
that fit this explanation. The high redshift ($z = 2.3$), extremely
metal-poor Damped Lyman-$\alpha$ (DLA) system recently reported by 
\citet[][{[Fe/H]} $\sim -3.0$]{cooke11} 
exhibits an enhanced carbon-to-iron ratio ([C/Fe] $= +1.5$) and
other elemental abundance signatures that \citet{kobayashi11} associate with
production by faint supernovae which experienced mixing and fallback. 

Another possible progenitor for this class is massive, rapidly rotating,
mega metal-poor ([Fe/H] $< -6.0$) stars, which models suggest have
greatly enhanced abundances of CNO due to distinctive internal burning
and mixing episodes, followed by strong mass loss \citep{hirschi06,
meynet06,meynet10a,meynet10b}. It is also of interest that
\citet{matsuoka11} have reported evidence for strong carbon production
in the early universe, based on their analysis of the optical spectrum
of the most distant known radio galaxy, TN J0924-2201, with $z = 5.19$.
There is the exciting possibility that the same or similar progenitors,
likely associated with the very first generations of stars, are
responsible for the distinctive abundance patterns in both the CEMP-no
stars and the high-redshift gas and galaxies.

{\scriptsize
\begin{table*}
\caption{Stars with anomalously high abundances of O, Na, Mg, Al, or Si compared
with Ca. For elements other than Ca, there are many values around or above
[X/Fe] = +1.4.} 
\label{hiab} 
\centering
\setlength\tabcolsep{3.5pt}
\begin{tabular}{lrrrrrrrrrcll}
\noalign{\smallskip}
\hline \hline
 Star & [Fe/H] && [C/Fe] & [O/Fe] & [Na/Fe] & [Mg/Fe] & [Al/Fe] & [Si/Fe] & [Ca/Fe] && Cls. & Refs. \\
\hline
\noalign{\smallskip}
HE 1012-1540 & -3.43 && 2.22 &  2.25	    & 1.21	   & 1.88        &  0.93 &  1.07 \vline & 0.57 &&  no & 40 \\
HE 1327-2326 & -5.96 && 3.78 &  3.42	    & 2.73	   & 1.97        &  1.46 &  ...  \vline & 0.44 &&  li & 35, 39\\
CS 29498-043 & -3.54 && 2.09 &  2.43	    & 1.47	   & 1.75        &  0.27 &  0.82 \vline & 0.16 &&  no & 25,11,12\\
CS 22949-037 & -3.84 && 1.06 &  1.97	    & 1.77	   & 1.42        & -0.04 &  0.77 \vline & 0.37 &&  no & 6, 14, 15, 40\\
\cline{8-9}
CS 29528-028 & -2.86 && 2.77 &  ...	    & 2.68	   & 1.69 \vline &  ...  &  ...         & 0.46 &&  b+ & 37 \\
HE 1447+0102 & -2.47 && 2.48 &  ...	    & 1.07	   & 1.43 \vline &  ...  &  ...         & 0.88 &&  b+ & 37 \\
SDSS 1707+58 & -2.49 && 2.1  &  ...	    & 2.71	   & 1.13 \vline &  ...  &  ...         & 0.79 &&  b+ & 41 \\
CS 29497-034 & -2.90 && 2.68 &  ...	    & 1.48	   & 1.02 \vline & -0.01 &  ...         & 0.48 &&  rs & 5, 29, 37\\
\cline{7-7}
CS 22942-019 & -2.60 && 2.1  &  0.97	    & 1.44 \vline  & 0.65	 & -0.35 &  ...         & 0.68 &&  s  & 8, 12, 46\\
CS 22958-042 & -2.85 && 3.15 &  1.35	    & 2.85 \vline  & 0.32	 & -0.85 &  0.15        & 0.36 &&  no & 33 \\
CS 29528-041 & -3.30 && 1.59 &  1.40	    & 1.20 \vline  & 0.40	 & -0.85 & -0.20        & 0.40 &&  b+ & 33 \\
\cline{6-6}
CS 29497-030 & -2.67 && 2.58 &  1.58 \vline & 0.55	   & 0.49        & -0.67 & -0.01        & 0.50 &&  s  & 21, 27\\
CS 31080-095 & -2.85 && 2.69 &  2.35 \vline &-0.28	   & 0.65        & -0.95 &  0.05        & 0.17 &&  b- & 33 \\
HE 0557-4840 & -4.75 && 1.65 &$<$3.09\vline &-0.16	   & 0.25        & -0.61 &  ...         & 0.25 &&  li & 38 \\
\hline
CS 30314-067 & -2.85 && 0.5  &  ...	    &-0.08         & 0.42        & -0.10 &  0.80        & 0.22 &&  no & 10 \\
CS 29502-092 & -2.76 && 1.0  &  ...	    &-0.01         & 0.37        & -0.77 &  0.84        & 0.27 &&  no & 10 \\
\noalign{\smallskip}
\hline
\end{tabular}

\tablefoot{The tenth column shows the classification of the CEMP stars according to \citet{beers05}: 
s = CEMP-s, rs = CEMP-rs, no = CEMP-no, b+ = stars with [Ba/Fe] $>$ 1 with no Eu abundance, 
b- = stars with 0 $<$ [Ba/Fe] $<$ 1 with no Eu abundance, li = stars with only an upper 
limit for both the Eu and Ba abundances.
The last column shows the abundance references, as enumerated below. For stars with more than 
one reference, the adopted value is the average or one of the references, according to 
criteria explained in the text.
}

\tablebib{
(5) \citet{hill00}; (6) \citet{norris01}; (8) \citet{preston01}; (10) \citet{aoki02a}; (11) \citet{aoki02b}; 
(12) \citet{aoki02c}; (14) \citet{depagne02}; (15) \citet{norris02}; (21) \citet{siva04}; (25) \citet{aoki04}; 
(27) \citet{ivans05}; (29) \citet{barb05}; (33) \citet{siva06}; (35) \citet{aoki06b}; (37) \citet{aoki07}; 
(38) \citet{norris07}; (39) \citet{frebel08}; (40) \citet{cohen08}; (41) \citet{aoki08}; (46) \citet{masseron10}.
}
\end{table*}
}

\subsection{Neutron-capture elements} 

\subsubsection{Abundances} 

\begin{figure*}
\centering
\includegraphics[width=16cm]{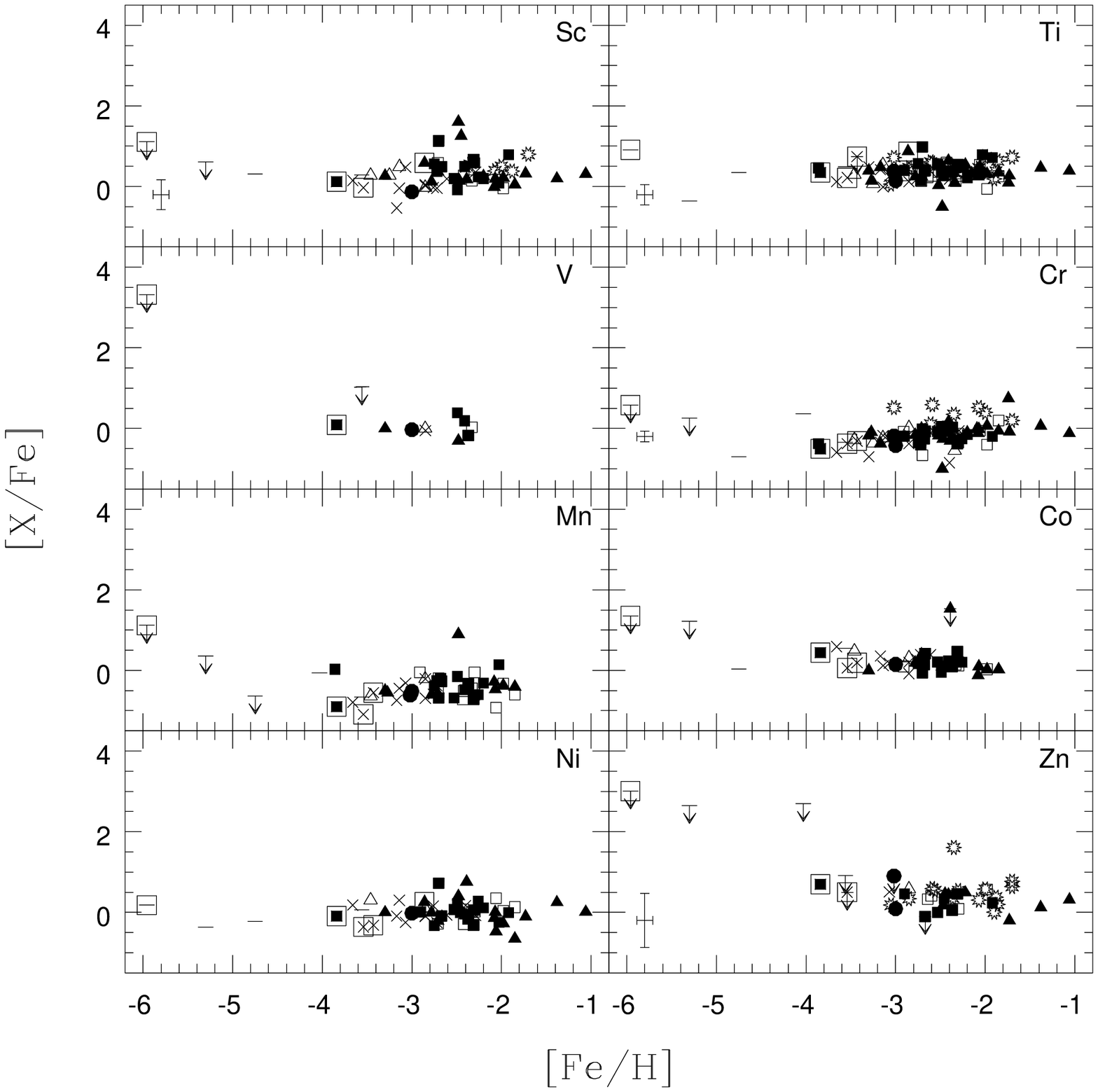}
\caption{[X/Fe] for CEMP stars of this work and from the literature. 
Symbols are the same as in Fig. \ref{15615fg1}}
\label{15615fg2}
\end{figure*}

\begin{figure*}
\centering
\includegraphics[width=16cm]{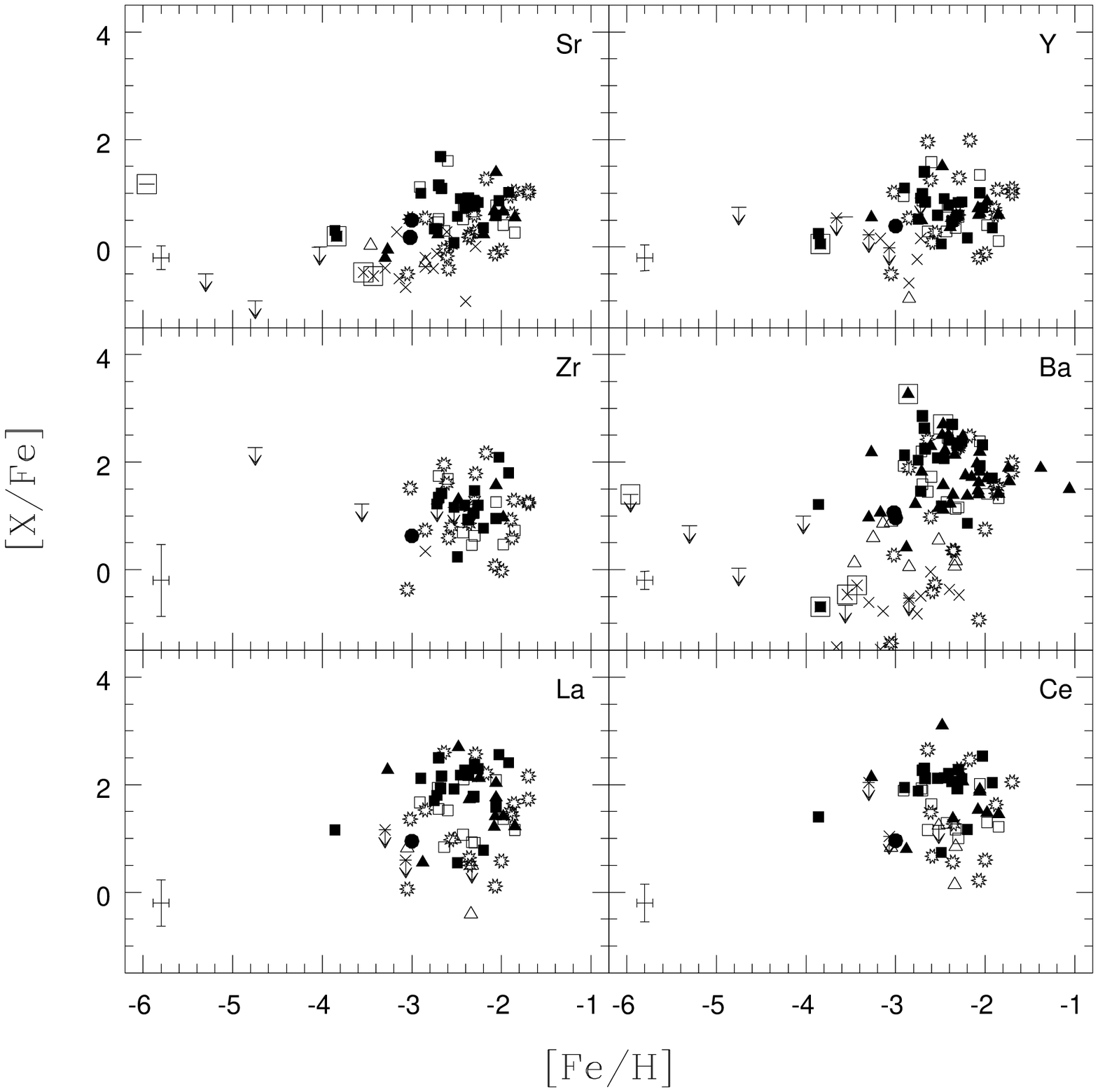}
\caption{[X/Fe] for CEMP stars of this work and from the literature. 
Symbols are the same as in Fig. \ref{15615fg1}}
\label{15615fg3}
\end{figure*}

\begin{figure*}
\centering
\includegraphics[width=16cm]{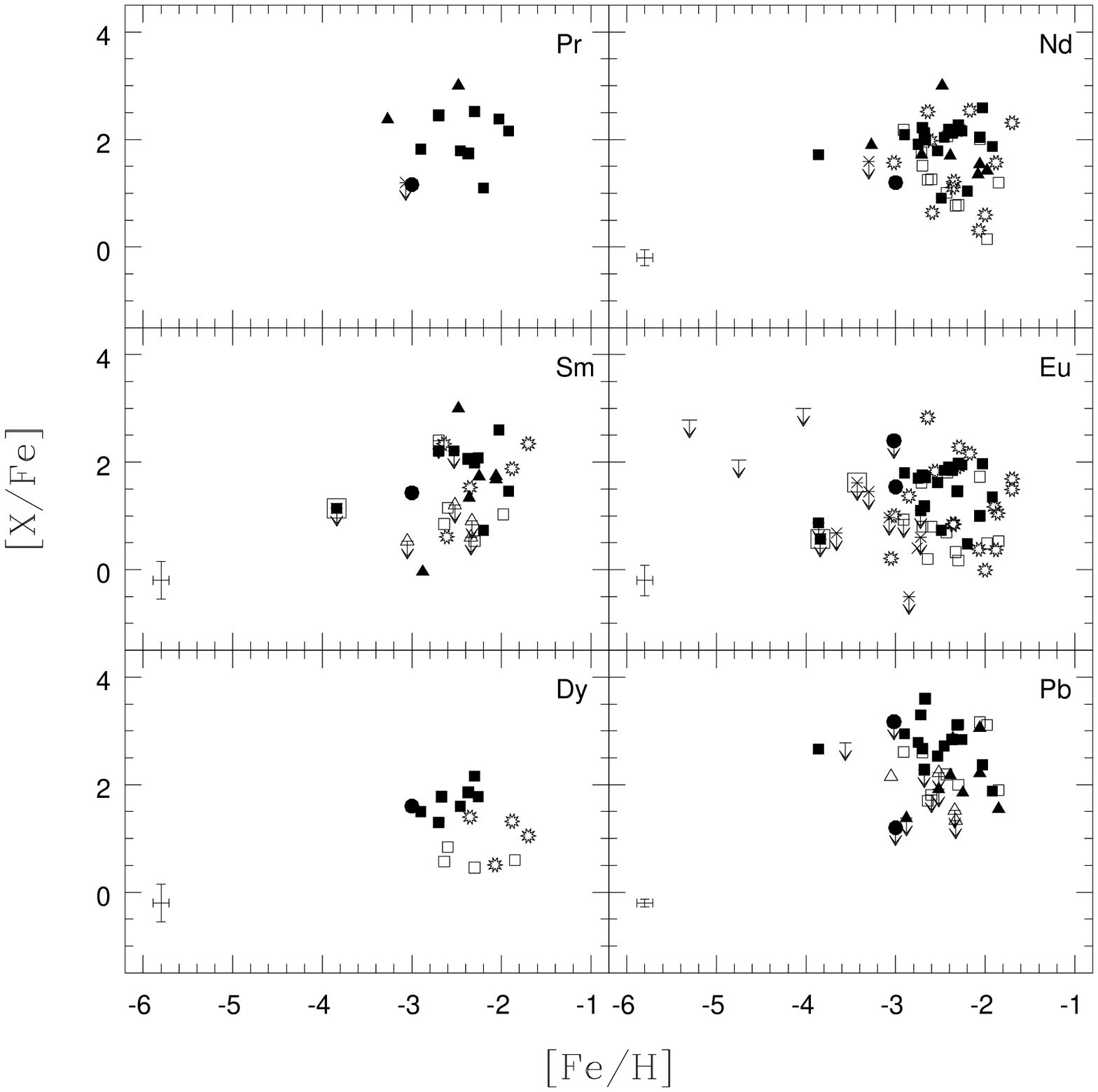}
\caption{[X/Fe] for CEMP stars of this work and from the literature. 
Symbols are the same as in Fig. \ref{15615fg1}}
\label{15615fg4}
\end{figure*}

\begin{figure*}
\centering
\includegraphics[width=16cm]{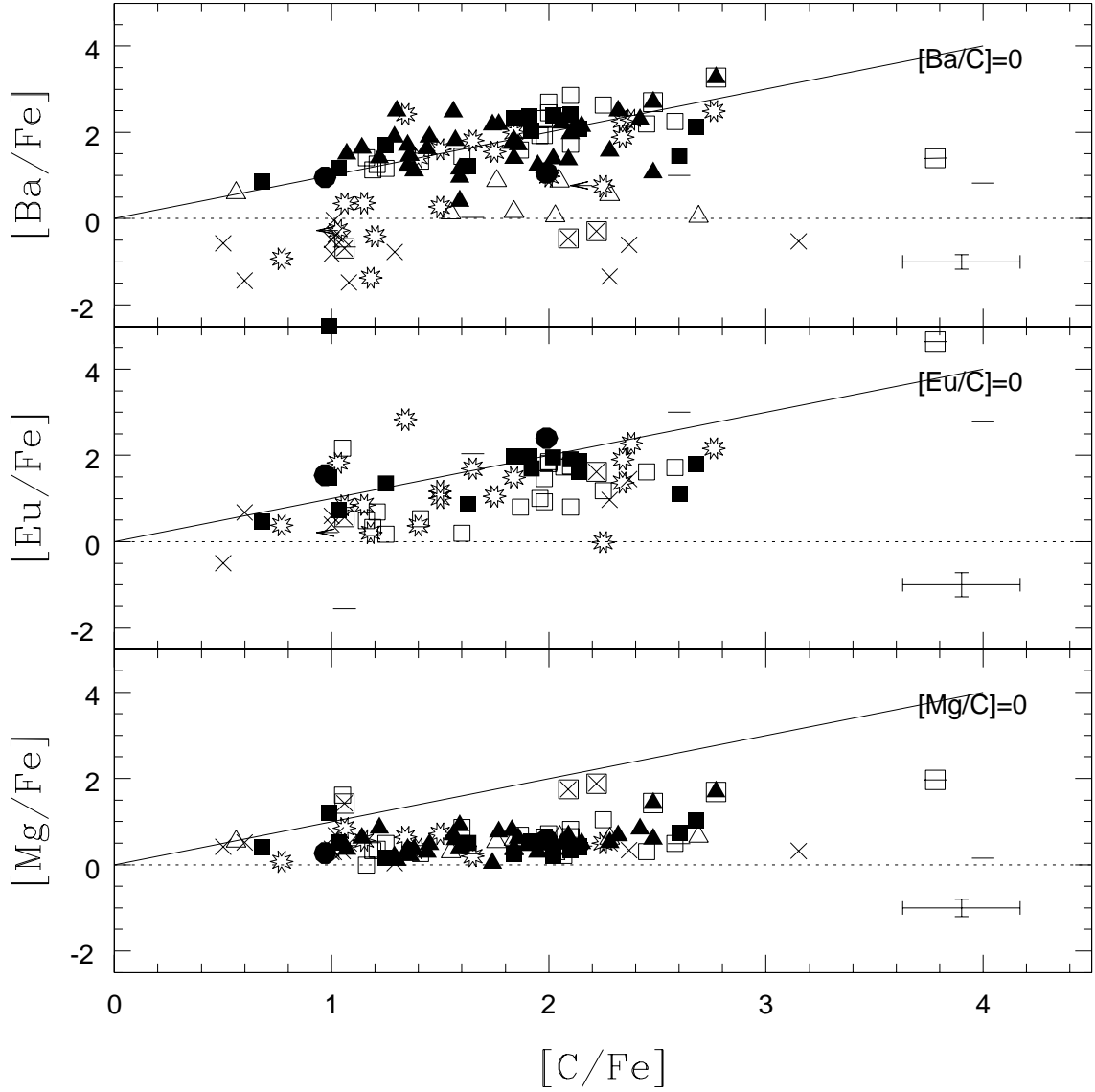}
\caption{[Mg,Eu,Ba/Fe] vs. [C/Fe] for CEMP stars of this work and from the literature. 
Fiducial lines [Ba, Eu, Mg/C] = 0 are drawn for convenience.
Symbols are the same as in Fig. \ref{15615fg1}. The arrows toward left indicate upper limit 
for [C/Fe].}
\label{15615fg5}
\end{figure*}

The behaviour of the neutron-capture-element abundances shown in 
Figs. \ref{15615fg3} and \ref{15615fg4} is remarkably different from the 
behaviour seen in Figs. \ref{15615fg1} and \ref{15615fg2} for the lighter 
elements. While in Figs. \ref{15615fg1} and \ref{15615fg2} all symbols are 
mixed together, the scatter is much
higher in Figs. \ref{15615fg3} and \ref{15615fg4}. However, if one
considers only data with the same symbol (excluding data of this work),
the scatter is lower than when considering all data. Given that the different
symbols are defined according to Ba and Eu abundances, this separation
is most likely a consequence of the contribution of the r- and
s-processes, which increase the abundances of neutron-capture elements
in different degrees for each class of star. It is worth recalling that
the difference among the procedures adopted by different authors might
increase the scatter observed in the reported abundance patterns.

The pattern of abundances of the neutron-capture elements for the stars
with large over-abundances of Mg (large open squares involving other
symbols) is not as clear as for the lighter elements. The stars
CS~22949-037, CS~29498-043, and HE~1012-1540 have [Ba/Fe] $<$ 0, whereas
CS~29528-028, HE~1447+0102 and HE~1327-2326 have very high barium
abundances, with [Ba/Fe] = +3.27 (+2.49 in this work\footnote{The origin
of the large difference between the Ba abundance of this star obtained
in this work and \citet{aoki07} might be the significant difference in
atmospheric parameters (see Table \ref{compar}) adopted by these
works.} ), +2.70, and +1.40, respectively. Three of these stars,
CS~22949-037, HE~1012-1540, and HE~1327-2326 have only upper limits for
[Eu/Fe] published ($<$+0.57, $<$+1.62, $<$+4.64, respectively), and the
three remaining have no published Eu abundance. The lack
of accurate abundances for [Eu/Fe] prevents a more meaningful analysis
of the possible relationship between the $\alpha$-elements and the
r-process elements. 

Figure \ref{15615fg5} exhibits an increasing trend for [Ba/Fe] vs.
[C/Fe]. While not all C-enhanced stars are Ba enhanced, those that are
lie on or below the envelope at [Ba/C] = 0; a fiducial line with [Ba/C]
= 0 is drawn to guide the eye. This behaviour of Ba --- a spread in
[Ba/C] with an upper bound at constant [Ba/C] (where C is presumably
$^{12}$C) --- might be understood as the consequence of the primary
production of $^{12}$C in He burning and up to but not exceeding an
equilibrium amount of $^{13}$C (the latter in hydrogen-ingestion via
$^{12}$C(p,$\gamma$)$^{13}$N($\beta^+$$\nu_{\rm e}$)$^{13}$C), and the
production of Ba proportionately to $^{13}$C through the
$^{13}$C($\alpha$,n)$^{16}$O source, in a TP-AGB star whose wind is then
transferred to its lower-mass companion forming the CEMP-s and CEMP-r/s
binary stars. 

As noted by \citet{tsang05}, [Eu/Fe] vs. [C/Fe] exhibits a similar
increasing trend (though many stars are missing in this panel compared
to the upper one for Ba because they have no Eu abundance available).
The upper envelope to [Eu/C] may be interpreted in the same way as the
maximal s-process Eu production, alongside Ba, for a given level of
$^{12}$C. There are also many stars known with r-process
enhancements but no C-enhancement; the origin of genuine r-process
enhancements might well be independent of the process responsible for
carbon and nitrogen excesses \citep[e.g.,][]{aoki02a,hansen11}.

\subsubsection{An upper limit on [C/H]}

In order for this scenario to have merit, we require primary production
of C in TP-AGB stars, and require this signature to remain even after
transfer onto a lower mass companion in an environmental CEMP-star
production event. Such evidence already exists, as it has been noted
already by \citet{ryan03} that the [C/Fe] vs [Fe/H] relation of CEMP
stars likewise exhibits an upper envelope consistent with [C/H] =
constant, close to zero. This implies that TP-AGB stars may produce up
to, but not in excess of, this level of C irrespective of metallicity
across the full metallicity range of halo stars. Precisely this
behaviour is seen in the yields of TP-AGB models of \citet{karakas02}.
Specifically, Karakas' carbon yields show that the ejected carbon mass,
while strongly dependent on progenitor mass --- minimal for
stars below 1.5~M$_\odot$ and peaking for stars around 3 to 4~M$_\odot$ ---
are almost independent of metallicity. Karakas presents calculations for
$Z$ = 0.020, 0.008, 0.004, i.e. over a range of a factor of five, and
shows that while the progenitor mass at which the ejected carbon mass
peaks does shift slightly -- but most importantly -- the level does not
change markedly; in other words ejected carbon is produced as a primary
element independent of metallicity. Although these AGB models have a
higher metallicity than genuine halo stars, the lack of metallicity
sensitivity over this broad range makes it plausible that the carbon is
produced in a primary fashion even as metallicity reduces further. In
the environmental enrichment scenarios for the production of CEMP stars,
the progenitor-mass dependence of ejected carbon mass and the various
accretion geometries will ensure that recipient stars achieve a range of
accreted carbon masses, and hence a range of [C/Fe] values, but the
metallicity independence of the ejected carbon mass, and the presumed
metallicity independence of the accretion mass, will ensure there is no
metallicity dependence of [C/H], thus providing an explanation for the
upper envelope for [C/H] seen in the population of CEMP
stars. A prediction coming from this interpretation, which may be
testable given larger samples in the future, is that CEMP stars formed
by other, non-AGB, mechanisms need not be restricted by this limit.

\subsubsection{Classification of CEMP-s stars} 

\begin{figure*}
\centering
\includegraphics[width=16cm]{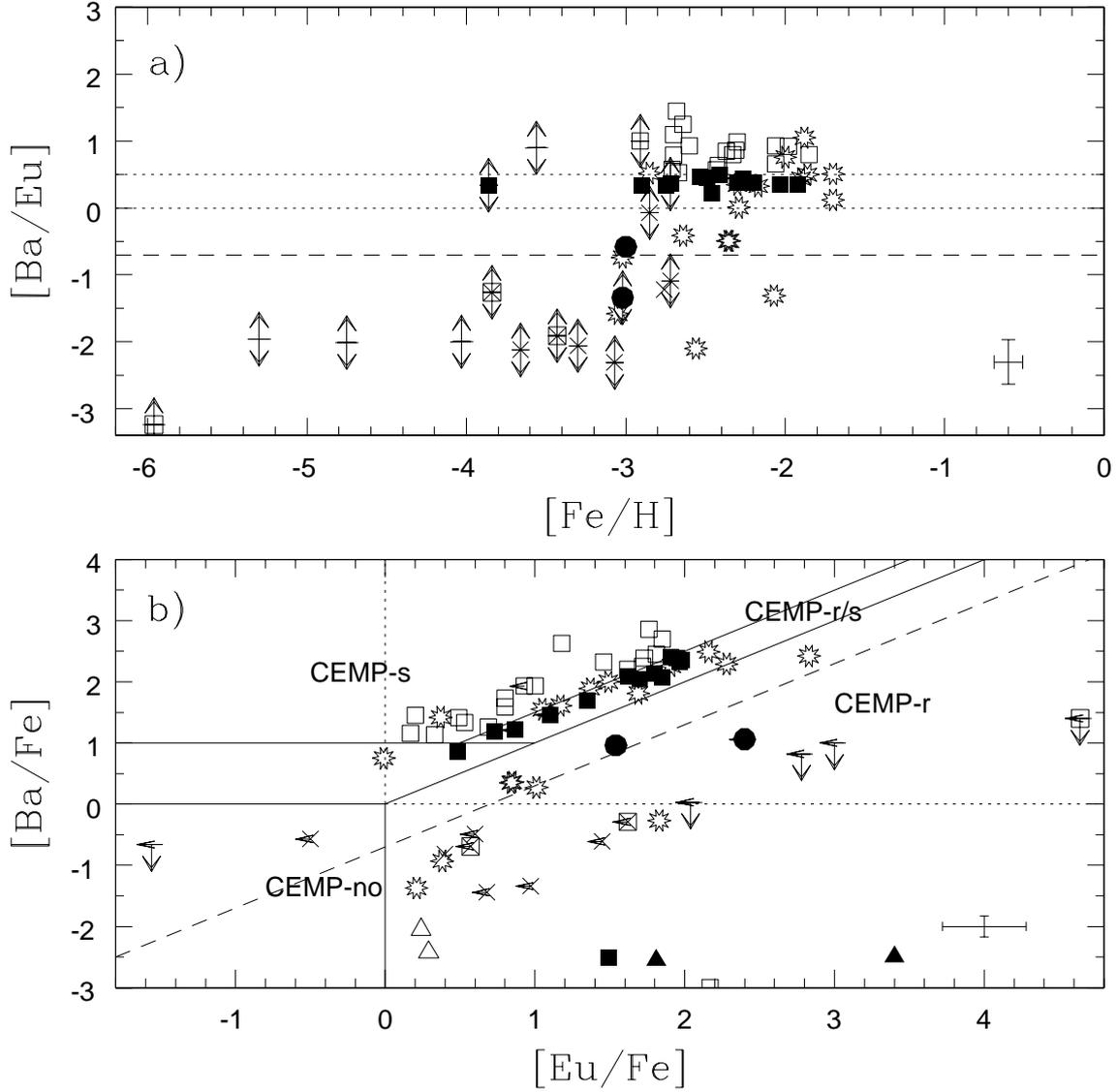}
\caption{CEMP stars from the literature and this work with the same symbols as in Fig. \ref{15615fg1}.
a) [Ba/Eu] vs. [Fe/H]. The dotted lines are the limits of the 
range for CEMP-r/s and the dashed line is the limit for the r-process only, [Ba/Eu] = $-$0.7.
The arrows up and down indicate upper limit for [Eu/Fe] and [Ba/Fe], respectively. 
b) [Ba/Fe] vs. [Eu/Fe]. Dashed line indicates [Ba/Eu] = $-$0.7, and 
dotted lines indicate where [Ba/Fe] = 0 and [Eu/Fe] = 0; 
CEMP-s and CEMP-r/s share instinctively the limits [Ba/Fe] $>$ 0 and [Ba/Fe] = [Eu/Fe],
whereas CEMP-no are inside the limits [Ba/Fe] and [Eu/Fe] $<$ 0 and 
CEMP-r inside [Eu/Fe] $>$ 0 and [Ba/Fe] = [Eu/Fe]. The arrows toward left and down indicate upper 
limit for [Eu/Fe] and [Ba/Fe], respectively.}
\label{15615fg7}
\end{figure*}

Figures \ref{15615fg7}a and \ref{15615fg7}b show the relationship
between Fe, Ba and Eu. Fig. \ref{15615fg7}b indicates the
conventional CEMP classification boundaries. CEMP stars exhibiting large
over-abundances of the s-process elements, with [Ba/Fe] $>$ +1 and
[Ba/Eu] $>$ +0.5, are CEMP-s, and are represented by the open squares in
Figs. \ref{15615fg1} to \ref{15615fg4}. The environmental scenario
suggested for their neutron-capture element enrichment involves mass
transfer in a binary system. The initially more massive star evolves
first and enters the TP-AGB phase, where it convectively enriches its
atmosphere with s-process products. After a phase of strong mass loss
through a stellar wind, it becomes a white dwarf, and now the former
secondary presents in its atmosphere vestiges of the nucleosynthesis of
the previous TP-AGB star. Nineteen stars from the literature were found
in our search to be in this class. Two stars of our sample (BS~16077-077
and CS~29512-073) might be CEMP-s (Fig. \ref{15615fg7}a). Although 
the Ba abundance of BS~16077-077 is not so high
([Ba/Fe] = +0.75), the low Eu abundance ([Eu/Fe] = $-0.01$) increases
the ratio [Ba/Eu]. This star is close to the line [Eu/Fe] $\sim$ 0 in
panel (b). 

\subsubsection{Classification of CEMP-r and CEMP-no stars} 

The CEMP stars that exhibit [Ba/Eu] $<$ 0 are referred to as CEMP-r. As
an example, the mean values for the star CS~22892-052 \citep{mcwill95,
norris97, sneden03,honda04,barklem05} are: [Eu/Fe] = +1.54 and [Ba/Fe] =
+0.96, resulting in [Ba/Eu] $= -0.59$ (see Tables \ref{heavy1} and
\ref{heavy2}). \citet{cohen06} reported a very high value of the Eu
abundance for the star HE~1410-0004, so this star is likely the second
CEMP-r star found in the literature. It is worth noting, however, that
their high Eu abundance is only an upper limit ([Eu/Fe] $<$ +2.40), and
therefore, only a lower limit in [Ba/Eu] $> -$1.34, can be inferred.
These two stars are represented by circles in the figures. 

Results from \citet{hansen11} support the hypothesis that r-process
enhancement is not coupled with the presence of a binary
companion, and suggests that where it is found to occur, it must have
originated in separate pollution events of the natal molecular clouds
from which CEMP-r stars formed. It is worth
noting that CS~22892-052 is included in the sample of r-process enhanced
stars that received extensive radial-velocity monitoring in the work of
\citet{hansen11}. They report that this star appears {\it not to be
a binary}, despite the earlier suggestion by \citet{preston01},
indicating that its C content was not produced by a companion. Thus,
membership in a binary system appears to be decoupled from details
associated with these particular abundance variations.

The CEMP stars with [Ba/Fe] $<$ 0 were called CEMP-no by \citet{ryan05}.
However, some of the stars in the literature with [Ba/Fe] $<$ 0 might be
CEMP-r, rather than CEMP-no, if [Eu/Fe] $>$ +1. Since some of them have
no Eu abundance available, they could not be classified accordingly. 

Five stars of this work have [Ba/Eu] $<$ 0, the region of CEMP-no and
CEMP-r, where two of them are below the dashed line in Fig. \ref{15615fg7}a, 
which represents the production of Ba only through the
r-process, [Ba/Eu] $\sim -0.7$ \citep{mg01}. One of them (BS~16033-081),
might be called CEMP-no, since its [Ba/Fe] $<$ 0, whereas the other
(BS~16929-005) might be CEMP-r, since [Ba/Fe] $\sim$ 0
\footnote{\citet{honda04} and \citet{aoki07} determined a very low value of [Ba/Fe] for
BS~16929-005, as can be seen in Table \ref{compar}.} and [Eu/Fe] $>$ +1.
If no Eu abundance is given, one may classify this star as CEMP-no. Two
stars with [Ba/Fe] $\sim$ +0.35 (BS~15621-047, BS~16543-097) might be
CEMP-r, since [Ba/Eu] $<$ 0 for both and [Eu/Fe] $\sim$ +1. The last
star, CS~31070-073, has the highest value of [Eu/Fe], even higher than
the ones from the literature referred to as CEMP-r (see filled circles
in Fig. \ref{15615fg7}b). [Ba/Eu] is very low, even though [Eu/Fe] and
[Ba/Fe] are high, suggesting that it might be CEMP-r/s.

\subsubsection{Classification of CEMP-r/s stars} \label{Crs}

For CEMP stars with 0 $<$ [Ba/Eu] $\lesssim$ +0.5, both r-process and
s-process enhancements have been inferred \citep{hill00,wanajo06}.
Fifteen CEMP stars corresponding to this description were found in the
literature so far, and are referred to as CEMP-r/s (or variations).
\citet{cohen03} was the first group to propose that the origin of
CEMP-r/s stars includes two mass transfers -- the first one would result
in the enrichment of s-process (from an AGB member of a binary system),
while the second would involve a so-called accretion-induced collapse
(AIC), as proposed by \citet{qian03}, at which time the r-process
enrichment would occur. After those events, the survivor star would be
observed as a CEMP-r/s star. \citet{qian01} commented on the possibility
that stars like CS~31082-001, which is r-process rich (but not C-rich),
has survived the SN II explosion of its companion and now has a compact
companion, most likely a black hole.

Three stars in the present work have ratios in the range 0 $<$ [Ba/Eu]
$<$ +0.5, the CEMP-r/s zone represented in Fig. \ref{15615fg7}b:
CS~29503-010, CS~29526-110, and CS~29528-028. One of them, CS~29526-110,
was similarly classified as CEMP-r/s by \citet{aoki02d}, but the other
two are being classified for the first time, since Aoki and
collaborators did not determine their Eu abundances (see Table
\ref{compar}).

\citet{ivans05} have derived abundances for CS~29497-030, and 
compared their results to theoretical models of neutron-capture
processes at low metallicity, both with and without pre-enrichment of
the initial abundances. In the pre-enrichment in r-elements case, the
heavy-element isotopes act as strong poisons for the neutron exposure,
competing with Fe and decreasing s-process efficiency.  

\citet{ivans05} claim that CS~29497-030 could be classified as 
CEMP-r/s, since their abundance ratios are best fitted by a
pre-enrichment of r-process material out of which the binary system
formed. However, \citet{ivans05} derived values 0.15 dex and 0.55 dex
higher than \citet{siva04} for [Ba/Fe] and [Eu/Fe], respectively. The
\citeauthor{siva04} [Eu/Fe] is better fitted by the model without
pre-enrichment in Fig. 2 of \citeauthor{ivans05} Therefore,
\citet{siva04} would classify as CEMP-s; the same situation pertains if
the classification comes from the average of the two works, as shown in
Table \ref{alpha}.

Although the CEMP-r/s classification has been widely used, we suggest 
\citep[following][]{tsang05} that it is a superfluous class, and that
there is no physical distinction between CEMP-s and CEMP-r/s stars,
other than the arbitrary choice of the [Ba/Eu] boundary. The r/s
classification was motivated by a belief that the high [Eu/Fe] values
and intermediate [Ba/Eu] values must signify Eu production by the
r-process in a site distinct from the one where Ba was produced via the
s-process. However, CEMP-s stars are able to produce quite high [Eu/Fe]
values, easily up to +2.0 (see Table \ref{heavy2}). The reasonable
conclusion is that Eu production, although r-process dominated in the
Solar System, is readily achievable by the s-process alone. Three
further arguments strengthen this view. First, if Eu production and Ba
production in CEMP-r/s stars is genuinely decoupled into separate r- and
s-processes, it is astonishing how remarkably tight the correlation
between [Ba/Fe] and [Eu/Fe] for CEMP-s and CEMP-r/s stars appears in
Fig. \ref{15615fg7}b. Two separate processes would naturally be
expected to provide more diverse outcomes. Secondly, Fig. \ref{15615fg5}
shows that the upper envelope of [Eu/Fe] in CEMP stars closely
corresponds to [Eu/C] = 0. That is, irrespective of the degree of the C
over-abundance in CEMP stars, Eu production is coupled to that value, at
least as an upper bound. The r-process, occurring presumably in the
final stage of evolution of these stars' progenitors, should not be
expected to follow the C abundance so closely. Thirdly, we note that
high Eu abundances are found in Pop. I Ba stars, where the r-process is
not believed to be active at all \citep{di06}. For the CEMP stars, a
more reasonable view (and certainly the view more consistent with
Occam's razor) is that {\it only one process} is responsible for the
production of both Ba and Eu in lockstep with C, and that process is the
s-process, not a tuned balance between s- and r-processes under widely
different physical conditions and at different phases of stellar
evolution. 

\subsubsection{Eu detections} \label{Eu}

\begin{figure*}
\centering
\includegraphics[width=16cm]{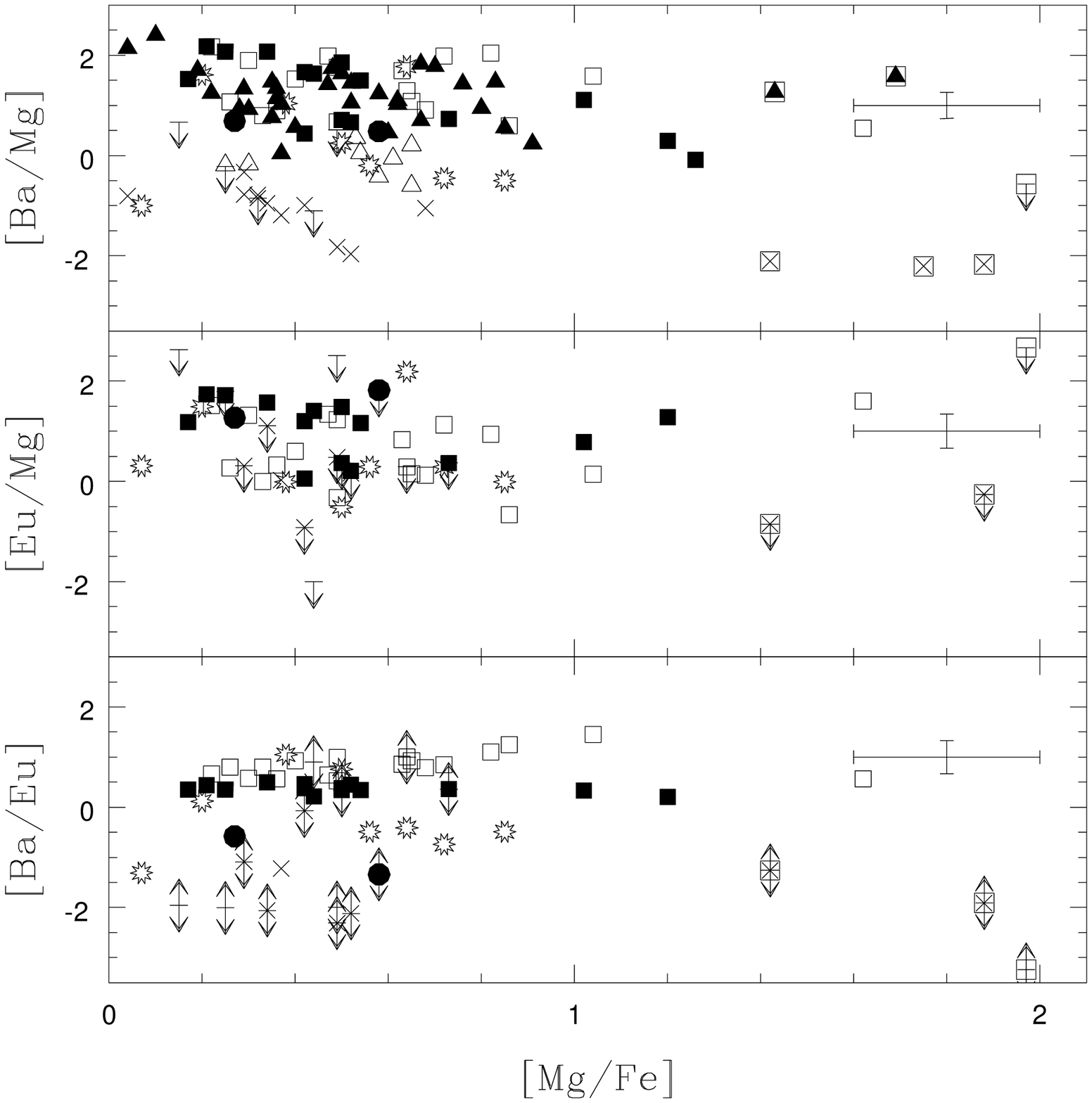}
\caption{The behaviour of Mg, Ba and Eu abundances for CEMP stars of this work and from the literature. 
Symbols are the same as in Fig. \ref{15615fg1}}
\label{15615fg6}
\end{figure*}

Our search for data from the literature reveals that the nucleosynthetic
record is seriously incomplete, particularly with respect to elements
associated with the r-process. About 67\% of CEMP stars have no
measured Eu abundance (or only an upper limit); this makes the abundance
pattern of many stars unclear. The lack of Eu abundances becomes clear
in Fig. \ref{15615fg6} through the open and filled triangles, which
represent stars for which there is no published abundance for Eu. The
missing crosses from one panel to another in Fig. \ref{15615fg6} 
also illustrate the lack of Eu for CEMP stars with [Ba/Fe] $<$ 0.
For some stars, only an upper limit was published either for Eu and Ba;
these stars are represented by dashes in Figs. \ref{15615fg1} to
\ref{15615fg7}. This work has provided Eu abundances for fifteen stars,
only four of which had a previous determination (see Table
\ref{compar}). For two stars of our sample it was not possible to
determine a Eu abundance, and for one, only an upper limit could be
estimated. \citet{masseron10} compiled abundances from analyses of
high-resolution spectra of metal-poor stars, including C-rich stars.
However, their suggestions of classifications for CEMP stars was
hampered due to the absence of Eu detection for many stars. The
difficulty in determining Eu abundances resides in the fact that its
lines in CEMP stars are very weak, so higher-quality spectra than have
been obtained to date are required (which in turn require long exposure times on large
telescopes). So, more research programs dedicated to the investigation
of Eu in CEMP stars, using new high-resolution
spectrographs on large telescopes, are needed to make further progress in
understanding the nucleosynthetic processes that took place in the early
Galaxy.

%

\section{Conclusions}\label{concl}

We have presented a detailed analysis of the elemental abundances for a
sample of 18 CEMP stars, based on high-resolution spectroscopic data,
and together with data from the literature, examined the behaviour of
their abundance ratios. Some stars exhibit large over-abundances of O,
Na, Mg, and Si: 3 stars are Si-rich; these 3 plus an additional 4 stars
are Mg-rich; these 7 plus 4 additional stars are Na-rich. Ten stars,
including some of these 11 Na-rich stars for which oxygen abundances
were available, are O-rich. These results are in agreement with the
fallback hypothesis \citep{turatto98,zampieri03}. However, stronger
conclusions must await suitable data from which O abundances for all of
these stars can be derived. 

This work has shown that the lack of accurate Eu abundances has made the
abundance pattern of many CEMP stars quite unclear. Once we have better
available data, it will be possible to perform an accurate detailed
analysis, allowing an improvement in our understanding of the
astrophysical origins of the different classes of CEMP stars. Observing
lines of Eu and other elements usually associated with the r-process
would indicate the true extent of the r-process. Notwithstanding the
relative paucity of Eu data, we contend on the basis of available C, Ba,
and Eu abundances that the hybrid CEMP-r/s class is superfluous, and
arises because of incomplete knowledge of Eu production by the s-process
at low metallicity. As a result, stars labeled as CEMP-r/s have the same
origin as CEMP-s stars. On the other hand, the presence of r-process
elements CEMP-r stars may well be due to previous pollution of the
molecular clouds from which they formed.
   
%

\begin{acknowledgements}

DMA acknowledges the University of Hertfordshire, where part of this
work was developed. DMA also acknowledges a CAPES post-doctoral
fellowship n$^{\circ}$ BEX 3448/06-1, as well as FAPESP, for the
post-doctoral fellowship n$^{\circ}$ 2008/01265-0, and CNPq, for the post-doctoral
fellowship n$^{\circ}$ 150370/2012-1. We are gratefull to
Fiorella Castelli for her prompt help concerning the codes ATLAS9,
WIDTH9 and SYNTHE. TCB acknowledges partial funding of this work from
grants PHY 02-16783 and PHY 08-22648: Physics Frontier Center/Joint
Institute for Nuclear Astrophysics (JINA), awarded by the U.S. National
Science Foundation. SR thanks the partial support from FAPESP, CNPq, and
CAPES. We acknowledge the anonymous referee for help in improving the
error analysis.

\end{acknowledgements}

%

\Online


\begin{appendix} 
\section{Uncertainty calculations}\label{apenderr}

Uncertainties on abundances were evaluated by consideration of the
degree to which a variation of 1$\sigma$ in the atmospheric parameters
(including macroturbulence for the SAT subsample) and the S/N (only for
the DMA subsample) affects the output value of the synthesis routine.
Table \ref{errDMASAT} shows the systematic effects that estimated errors
in the adopted atmospheric parameters have on abundances
($\log\epsilon$(X)).

For the DMA subsample, the star BS 16543-097 was used as a reference to
calculate uncertainties, since it has a similar temperature as many of
the other stars. One line of each element was chosen to calculate the
effects that errors on atmospheric parameters have on our derived
abundances; these were then extended to all other lines of the same
element.

Extending the results on uncertainties computed for one line to the
other lines of the same element is not actually the best procedure.
Different lines (i.e., lines with different EPs, strengths, etc.) may have
different dependences on the stellar parameters. In fact, the ideal
procedure would be to compute all variations for all lines. However, it
is impractical using spectrum synthesis, due to the very long time that
it would take, since each variation requires the complete ``ritual'' of
the full spectrum synthesis. However, if one chooses a suitable line as
a proxy for the remaining lines, it is valid to estimate the variations
due the atmospheric and experimental parameters for this line, and to
assume that this variation is representative for all other lines of the
same element. The profile of this line must to be as well-defined as
possible -- not too strong, not too weak, in order to make the line
representative. 

To determine the uncertainties on abundances, we followed \citet{J02}; a 
portion of this procedure and formulae are rewritten below.
Then, the variance in $\log\epsilon$(X) can be written as

{\small
\begin{eqnarray}
\label{errlogEp} 
\sigma_{\log\epsilon}^2=
\sigma_{rand}^2+
{\left(\partial\log\epsilon \over\partial T\right)}^2\sigma_T^2+ 
{\left(\partial\log\epsilon \over\partial \log g\right)}^2\sigma_{\log g}^2 + 
{\left(\partial\log\epsilon \over\partial \xi\right)}^2\sigma_{\xi}^2 + \nonumber \\
2{\left[
{\left(\partial\log\epsilon \over\partial T\right)}{\left(\partial\log\epsilon \over\partial \log g\right)}\sigma_{T\log g}\right]} +
2{\left[
{\left(\partial\log\epsilon \over\partial \xi\right)}{\left(\partial\log\epsilon \over\partial \log g\right)}\sigma_{\log g \xi}
\right]} + \nonumber \\
2{\left[
{\left(\partial\log\epsilon \over\partial \xi\right)}{\left(\partial\log\epsilon \over\partial T\right)\sigma_{\xi T}}
\right]}
\end{eqnarray}
}

\noindent where $\sigma_{rand}^2$ is the difference between the determined abundance 
for one line found through the best fit and that found after adding
(S/N)$^{\rm -1}$ to it. The term (S/N)$^{\rm -1}$ comes from the inverse
relation between S/N and the difference of the observed fluxes and the
synthetic spectrum used to determine the abundance of a line. The S/N
ratios shown in Table \ref{temps} were measured in a window around
$\lambda$4500\AA. The partial derivatives ${(\partial\log\epsilon /
\partial T)}$, ${(\partial\log\epsilon / \partial \log g)}$, and
${(\partial\log\epsilon / \partial \xi)}$ were determined by varying
1$\sigma$ on the model atmosphere for the star BS~16543-097, as
explained above.

The covariances $\sigma_{T\log g}$, $\sigma_{\log g \xi}$, and
$\sigma_{\xi T}$ also were determined using the star BS~16543-097. We
picked 11 values for temperature from $-1\sigma$ to $+1\sigma$, and
created model atmospheres with several values of $\log g$ to force
ionization balance. Then, the covariance between $T$ and $\log g$ is

{\small
\begin{equation}
\sigma_{T\log g} = {1\over N}\sum_{i=1}^N (T_i-T_m)(\log g_i-\log g_m),
\end{equation}
}

\noindent where $T_m$ and $\log g_m$ are the adopted temperature
and $\log g$ for the star. The correlation coefficient is defined as

{\small
\begin{equation}
\rho(T,\log g)={\sigma_{T\log g} \over \sigma_T \sigma_{\log g}},
\end{equation}
}

\noindent where in this case, $\sigma_T$ and $\sigma_{\log g}$ are the standard deviation
for $T$ and $\log g$, taking into account the 11 values mentioned above.
We found $\sigma_{T\log g}$ = 3.06 and $\rho(T,\log g)$= 1.02,
indicating a strong correlation between $T$ and $\log g$. We performed
a similar calculation to determine $\sigma_{\log g \xi} = -$0.00062 and
correlation $\rho(\log g\xi) = -$0.8197, indicating that $\log g$
and $\xi$ are highly correlated, although less so than for $T$ and $\log
g$. Similarly to \citet{J02}, $\sigma_{\xi T}$ was found to be
negligible, and $\sigma_{T\log g}$, the strongest correlation.

Considering the ratio of two elements A and B, its error is

{\small
\begin{equation}
\sigma^2(A/B)={\sigma_A^2 + \sigma_B^2 -2\sigma_{A,B}},
\end{equation}
}
\noindent where the covariance between two abundances is given by

{\small
\begin{eqnarray}
\label{errAB} 
\sigma_{A,B}=
{\left(\partial\log\epsilon_A \over\partial T\right)}{\left(\partial\log\epsilon_B \over\partial T\right)}\sigma_T^2 +
{\left(\partial\log\epsilon_A \over\partial \log g\right)}{\left(\partial\log\epsilon_B \over\partial \log g\right)}\sigma_{\log g}^2 + \nonumber \\
{\left(\partial\log\epsilon_A \over\partial \xi\right)}{\left(\partial\log\epsilon_B \over\partial \xi\right)}\sigma_\xi^2 + \nonumber \\
{\left[
{\left(\partial\log\epsilon_A \over\partial T\right)}{\left(\partial\log\epsilon_B \over\partial \log g\right)} + 
{\left(\partial\log\epsilon_A \over\partial \log g\right)}{\left(\partial\log\epsilon_B \over\partial T\right)}
\right]}\sigma_{T\log g} + \nonumber \\
{\left[
{\left(\partial\log\epsilon_A \over\partial \xi\right)}{\left(\partial\log\epsilon_B \over\partial \log g\right)} +
{\left(\partial\log\epsilon_A \over\partial \log g\right)}{\left(\partial\log\epsilon_B \over\partial \xi\right)}
\right]}\sigma_{\log g \xi}
\end{eqnarray}
}

Table \ref{medDMA} includes uncertainties on DMA subsample, while the
last column on Table \ref{errDMASAT} shows the estimate uncertainties
for the SAT subsample.

\begin{table}
\caption{Diference between the output value of the synthesis program ($\log A_p$) 
and the output value with 1$\sigma$ of diference in temperature ($\Delta_T$), 
$\log g$ ($\Delta_{lg}$),
metallicity ($\Delta_{feh}$), 
velocity of microturbulent ($\Delta_{\xi}$), 
observational ($\Delta_{Obs}$), 
and macroturbulence $\Delta_{mu}$. 
The last column of SAT subsample is the typical uncertainties for each element.}
\label{errDMASAT} 
\centering
\begin{tabular}{rrrrrrr}
\noalign{\smallskip}
\hline \hline
\noalign{\smallskip}
\multicolumn{7}{c} {DMA subsample} \\
\hline
Species & $\Delta_T$ & $\Delta_{lg}$ & $\Delta_{feh}$ & $\Delta_{\xi}$ & 
$\Delta_{Obs}$ & \\
\hline
\noalign{\smallskip}
 C & 0.03 & 0.01 & 0.01 & 0.01 & 0.20  & \\
Na & 0.02 & 0.01 & 0.01 & 0.01 & 0.30  & \\
Mg & 0.01 & 0.01 & 0.01 & 0.01 & 0.09  & \\
Sc & 0.01 & 0.02 & 0.02 & 0.01 & 0.40  & \\
Ti & 0.02 & 0.01 & 0.02 & 0.01 & 0.12  & \\
Cr & 0.03 & 0.01 & 0.01 & 0.01 & 0.08  & \\
\hline
\end{tabular}
\end{table}

\section{Data from the literature}\label{tabslit}

{\scriptsize
\begin{center}
\setlength\tabcolsep{3pt}
\begin{longtable}{lcccccccccccccccccclr}
\caption{Abundances [X/Fe] of light elements for CEMP stars.} \label{alpha} \\
\hline\hline
Star & $T_{\rm eff}$ & $\log g$ & [Fe/H] & $\xi$ & A(Li) & C & N & O & Na & Mg & Al & Si & S & K & Ca & BM & EM & BE & Cls. & Ref. \\
 & K &  &  & km/s &  &  &  &  &  &  &  &  &  &  &  &  &  &  &  &  \\
\hline 
\endfirsthead

\multicolumn{21}{c}%
{{\bfseries \tablename\ \thetable{} -- continue}} \\
\hline\hline
Star & $T_{\rm eff}$ & $\log g$ & [Fe/H] & $\xi$ & A(Li) & C & N & O & Na & Mg & Al & Si & S & K & Ca & BM & EM & BE & Cls. & Ref. \\
 & K &  &  & km/s &  &  &  &  &  &  &  &  &  &  &  &  &  &  &  &  \\
\hline 

\endhead
\endlastfoot
BS 16929-005 & 5270 & 2.7  & -3.09 & 1.3  & ...  & 0.92 & ...  & ...  & ...  & 0.38 & -0.85 & 0.38 & ...  & ...    & 0.46 &	  &	  &	  &	& 22 \\
BS 16929-005 & 5250 & 2.8  & -3.17 & 2.3  & ...  & 1.08 &$<$2.39& ...  &-0.06 & 0.49 &  ...  & ...  & ...  & ...    & 0.13 &	  &	  &	  &	& 37 \\
BS 16929-005 & 5250 & 2.8  & -3.17 & 2.3  & ...  & 1.08 &$<$2.39& ...  &-0.06 & 0.49 & -0.85 & 0.38 & ...  & ...    & 0.13 & ...   &  ...  &  ...  &  no &adopted\\
\hline
BS 17435-532 & 5200 & 2.15 & -2.2  & 2.0  & 2.06 & 0.68 & ...  & 1.06 & 0.69 & 0.42 &  ...  & ...  & ...  & 0.44   & 0.43 & 0.44  &  0.06 &  0.38 &  rs & 43 \\
\hline 
CS 22166-016 & 5250 & 2.0  & -2.40 & 2.2  & ...  & 1.02 & ...  & ...  & 0.37 & 0.68 &  ...  & 0.22 & ...  & ...    & 0.50 & -1.05 &  ...  &  ...  &  no & 7 \\
\hline 
CS 22877-001 & 5100 & 2.2  & -2.72 & 2.0  &$<$1.2 & 1.0  & 0.0  & ...  &-0.24 & 0.29 & -0.72 & ...  & ...  & ...    & 0.42 &	  &	  &	  &	& 10 \\
CS 22877-001 & 5000 & 1.5  & -2.82 & 2.3  & ...  & 1.80 & ...  & ...  &-0.47 & 0.14 &  0.14 & 0.08 & ...  & ...    & 0.31 &	  &	  &	  &	& 7 \\
CS 22877-001 & 5100 & 2.2  & -2.72 & 2.0  &$<$1.2 & 1.0  & 0.0  & ...  &-0.24 & 0.29 & -0.72 & ...  & ...  & ...    & 0.42 & -0.78 &  0.31 & -1.09 &  no &adopted\\
\hline
\end{longtable}
\end{center}

\tablefoot{The Li abundance is A(Li) = $\log\epsilon$(Li) + 12.0.
The last 5 columns are: BM = [Ba/Mg]; EM = [Eu/Mg]; BE = [Ba/Eu]; Cls. = classification of CEMP stars according to \citet{beers05}: 
s = CEMP-s, r = CEMP-r, rs = CEMP-rs, no = CEMP-no, b+ = stars with [Ba/Fe] $>$ 1 with no Eu abundance, 
b- = stars with 0 $<$ [Ba/Fe] $<$ 1 with no Eu abundance, N = stars with no Ba and Eu abundances, li = stars with only an upper 
limit for both, Eu and Ba abundances;
Ref. = abundance references, as enumerated below. The word ``adopted'' for stars with more than 
one reference indicates that the adopted value is the average or one of the references above, according to 
criteria explained in the text. Some stars are found in the literature with different IDs: 
HE~2148-1247 = CS~22944-0068, HE~0058-0244 = CS~22183-0015, CS~31062-012 = LP 706-7, G77-61 = HE~0330+0148, 
CS~31062-050 = HE~0027-1221, CS~29497-034 = HE~0039-2635, CS~22957-027 = HE~2356-0410, CS~22948-027 = HE~2134-3940, 
CS~22949-037 = HE~2323-0256, CS~22942-019 = HE~0054-2542, CS~22892-052 =
HE~2214-1654, BS~17435-0532 = HKII 17435-00532.
}

\tablebib{
(1) \citet{kipper94}; (2) \citet{mcwill95}; (3) \citet{norris97}; (4) \citet{aoki00}; (5) \citet{hill00}; (6) \citet{norris01}; 
(7) \citet{giridhar01}; (8) \citet{preston01}; (9) \citet{aoki01}; (10) \citet{aoki02a}; (11) \citet{aoki02b}; (12) \citet{aoki02c}; 
(13) \citet{aoki02d}; (14) \citet{depagne02}; (15) \citet{norris02}; (16) \citet{christ02}; (17) \citet{lucatello03}; (18) \citet{sneden03}; 
(19) \citet{cohen03}; (20) \citet{vaneck03}; (21) \citet{siva04}; (22) \citet{honda04}; (23) \citet{christ04}; (24) \citet{johnson04}; 
(25) \citet{aoki04}; (26) \citet{barklem05}; (27) \citet{ivans05}; (28) \citet{plez05}; (29) \citet{barb05}; (30) \citet{goswami06}; 
(31) \citet{cohen06}; (32) \citet{jonsell06}; (33) \citet{siva06}; (34) \citet{aoki06a}; (35) \citet{aoki06b}; (36) \citet{frebel07}; 
(37) \citet{aoki07}; (38) \citet{norris07}; (39) \citet{frebel08}; (40) \citet{cohen08}; (41) \citet{aoki08}; (42) \citet{thompson08}; 
(43) \citet{roederer08}; (44) \citet{goswami10}; (45) \citet{behara10}; (46) \citet{masseron10}.
}



\begin{center}
\setlength\tabcolsep{3pt}
\begin{longtable}{lccccccccccr}
\caption{Abundances [X/Fe] of iron-peak elements for CEMP stars.} \label{ironpeak} \\

\hline\hline
Star & [Fe/H] & Sc & Ti & V & Cr & Mn & Co & Ni & Cu & Zn & Ref. \\
\hline 
\endfirsthead

\multicolumn{12}{c}%
{{\bfseries \tablename\ \thetable{} -- continue}} \\
\hline\hline
Star & [Fe/H] & Sc & Ti & V & Cr & Mn & Co & Ni & Cu & Zn & Ref. \\
\hline 

\endhead
\endlastfoot
BS 16929-005 & -3.09 & -0.53 &  0.42  &  ...  &  ...   &  -0.74 &   0.36 &  -0.09  &  ...  &   ...   & 22 \\
BS 16929-005 & -3.17 &  ...  &  0.44* &  ...  & -0.34  &   ...  &   ...  &   ...   &  ...  &   ...   & 37 \\
BS 16929-005 & -3.17 & -0.53 &  0.42  &  ...  & -0.34  &  -0.74 &   0.36 &  -0.09  &  ...  &   ...   &adopted\\
\hline 
BS 17435-532 & -2.2  &  0.20 &  0.22  &  ...  & -0.15  &  -0.31 &   ...  &   0.11  &  ...  &   ...   & 43 \\
\hline 
CS 22166-016 & -2.40 &  0.48 &  0.18  &  ...  & -0.85  &   ...  &   ...  &   0.17  &  ...  &   ...   & 7 \\
\hline 
CS 22877-001 & -2.72 & -0.04 &  0.15  &  ...  & -0.29  &  -0.38 &   0.39 &  -0.25  &  ...  &   ...   & 10 \\
CS 22877-001 & -2.82 & -0.13 &  0.16  &  ...  & -0.18  &   ...  &   0.35 &  -0.07  &  ...  &   ...   & 7 \\
CS 22877-001 & -2.72 & -0.04 &  0.15  &  ...  & -0.29  &  -0.38 &   0.39 &  -0.25  &  ...  &   ...   &adopted\\
\hline
\end{longtable}
\end{center}

\tablefoot{References are the same as in Table \ref{alpha}. Ti abundances are of \ion{Ti}{i}, 
and the symbol * indicates \ion{Ti}{ii}. See notes in Table \ref{alpha} for different IDs.
}

\begin{center}
\setlength\tabcolsep{1.5pt}
\begin{longtable}{lcccccccccccccccccccr}
\caption{Abundances [X/Fe] of heavy elements for CEMP stars.} \label{heavy1} \\

\hline\hline
Star & [Fe/H] & Ga & Ge & Sr & Y & Zr & Nb & Mo & Ru & Rh & Pd & Ag & Cd & Sn & Ba & La & Ce & Pr & Nd & Ref. \\
\hline 
\endfirsthead

\multicolumn{21}{c}%
{{\bfseries \tablename\ \thetable{} -- continue}} \\
\hline\hline
Star & [Fe/H] & Ga & Ge & Sr & Y & Zr & Nb & Mo & Ru & Rh & Pd & Ag & Cd & Sn & Ba & La & Ce & Pr & Nd & Ref. \\
\hline 

\endhead
\endlastfoot
BS 16929-005 & -3.09 &  ...  &  ...  &  0.28 &  0.16 &  ...  &  ...  &  ...  &  ...  &  ...  &  ...  &  ...  &  ...  &  ...  & -0.59 &  ...  &  ...  &  ...  &  ...  & 22 \\
BS 16929-005 & -3.17 &  ...  &  ...  &  ...  &  ...  &  ...  &  ...  &  ...  &  ...  &  ...  &  ...  &  ...  &  ...  &  ...  & -1.48 &  ...  &  ...  &  ...  &  ...  & 37 \\
BS 16929-005 & -3.17 &  ...  &  ...  &  0.28 &  0.16 &  ...  &  ...  &  ...  &  ...  &  ...  &  ...  &  ...  &  ...  &  ...  & -1.48 &  ...  &  ...  &  ...  &  ...  &adopted\\
\hline 
BS 17435-532 & -2.2  &  ...  &  ...  &  0.36 & 0.17  & 0.77  &  ...  &  ...  &  ...  &  ...  &  ...  &  ...  &  ...  &  ...  &  0.86 &  0.78 & 1.17  & 1.10  & 1.04  & 43 \\
\hline 
CS 22166-016 & -2.40 &  ...  &  ...  & -1.01 &  ...  &  ...  &  ...  &  ...  &  ...  &  ...  &  ...  &  ...  &  ...  &  ...  & -0.37 &  ...  &  ...  &  ...  &  ...  & 7 \\
\hline 
CS 22877-001 & -2.72 &  ...  &  ...  & -0.12 &  0.16 &  ...  &  ...  &  ...  &  ...  &  ...  &  ...  &  ...  &  ...  &  ...  & -0.49 &  ...  &  ...  &  ...  &  ...  & 10 \\
CS 22877-001 & -2.82 &  ...  &  ...  & -0.38 & -0.27 &  ...  &  ...  &  ...  &  ...  &  ...  &  ...  &  ...  &  ...  &  ...  & -0.52 &  ...  &  ...  &  ...  &  ...  & 7 \\
CS 22877-001 & -2.72 &  ...  &  ...  & -0.12 &  0.16 &  ...  &  ...  &  ...  &  ...  &  ...  &  ...  &  ...  &  ...  &  ...  & -0.49 &  ...  &  ...  &  ...  &  ...  &adopted\\
\hline
\end{longtable}
\end{center}

\tablefoot{References are the same as in Table \ref{alpha}. See notes in Table \ref{alpha} for different IDs.
}

\begin{center}
\setlength\tabcolsep{1.5pt}
\begin{longtable}{lcccccccccccccccccccccr}
\caption{Abundances [X/Fe] of heavy elements for CEMP stars.} \label{heavy2} \\

\hline\hline
Star & [Fe/H] & Sm & Eu & Gd & Tb & Dy & Ho & Er & Tm & Yb & Lu & Hf & W & Os & Ir & Pt & Au & Pb & Bi & Th & U & Ref. \\
\hline
\endfirsthead

\multicolumn{23}{c}%
{{\bfseries \tablename\ \thetable{} -- continue}} \\
\hline\hline
Star & [Fe/H] & Sm & Eu & Gd & Tb & Dy & Ho & Er & Tm & Yb & Lu & Hf & W & Os & Ir & Pt & Au & Pb & Bi & Th & U & Ref. \\
\hline

\endhead
\endlastfoot
BS 16929-005 & -3.09 &  ...  &  ...  & ...  & ...  & ...  & ...  & ...  & ...  & ...  & ...  & ...  &  ... &  ...  &  ...  &  ...  &  ...  &  ...  &  ...  & ...  &  ... & 22 \\
BS 16929-005 & -3.17 &  ...  &  ...  & ...  & ...  & ...  & ...  & ...  & ...  & ...  & ...  & ...  &  ... &  ...  &  ...  &  ...  &  ...  &  ...  &  ...  & ...  &  ... & 37 \\
BS 16929-005 & -3.17 &  ...  &  ...  & ...  & ...  & ...  & ...  & ...  & ...  & ...  & ...  & ...  &  ... &  ...  &  ...  &  ...  &  ...  &  ...  &  ...  & ...  &  ... &adopted\\
\hline 
BS 17435-532 & -2.2  & 0.73  &  0.48 & ...  & ...  & ...  & ...  & ...  & ...  & ...  & ...  & ...  &  ... &  ...  &  ...  &  ...  &  ...  &   ... &  ...  & ...  &  ... & 43 \\
\hline 
CS 22166-016 & -2.40 &  ...  &  ...  & ...  & ...  & ...  & ...  & ...  & ...  & ...  & ...  & ...  &  ... &  ...  &  ...  &  ...  &  ...  &  ...  &  ...  & ...  &  ... & 7 \\
\hline 
CS 22877-001 & -2.72 &  ...  & $<$0.6 & ...  & ...  & ...  & ...  & ...  & ...  & ...  & ...  & ...  &  ... &  ...  &  ...  &  ...  &  ...  &  ...  &  ...  & ...  &  ... & 10 \\
CS 22877-001 & -2.82 &  ...  &  ...  & ...  & ...  & ...  & ...  & ...  & ...  & ...  & ...  & ...  &  ... &  ...  &  ...  &  ...  &  ...  &  ...  &  ...  & ...  &  ... & 7 \\
CS 22877-001 & -2.72 &  ...  & $<$0.6 & ...  & ...  & ...  & ...  & ...  & ...  & ...  & ...  & ...  &  ... &  ...  &  ...  &  ...  &  ...  &  ...  &  ...  & ...  &  ... &adopted\\
\hline		       
\end{longtable}
\end{center}

\tablefoot{References are the same as in Table \ref{alpha}. See notes in Table \ref{alpha} for different IDs.
}

}

\end{appendix}

\end{document}